\documentclass[11pt, a4paper]{article}

\usepackage[a4paper, top=2.5cm, bottom=2.5cm, left=2cm, right=2cm]{geometry}
\usepackage[english]{babel}
\usepackage[T1]{fontenc}
\usepackage{newtxtext}

\usepackage{amsmath}
\usepackage{amssymb}
\usepackage{amsfonts}
\usepackage{bm}
\usepackage{microtype}
\usepackage{xcolor}
\usepackage{booktabs}

\usepackage{simplewick}

\usepackage{jheppub}

\newcommand{\p}{\partial}

\def\>{\rangle} \def\<{\langle}

\title{\boldmath Quantum Anomalies of Tensionless Bosonic Strings}

\author[a,b,c]{Bin Chen}
\author[b]{Zezhou Hu}

\affiliation[a]{Center for Gravitation and Astrophysics, Faculty of Science,\\
Kunming University of Science and Technology, Kunming 650500, China}
\affiliation[b]{School of Physics, Peking University, \\No.5 Yiheyuan Rd, Beijing 100871, P.~R.~China}
\affiliation[c]{Institute of Fundamental Physics and Quantum Technology, \\\&  School of Physical Science and Technology, \\ Ningbo University, Ningbo, Zhejiang 315211, China}

\emailAdd{chenbin1@nbu.edu.cn, z.z.hu@pku.edu.cn}

\abstract{
We systematically investigate and compare the worldsheet actions, BRST structures and the quantum anomalies of four different formulations of tensionless ($T = 0$) bosonic string theory: the $(D+2)$-dimensional conformal string \cite{Gustafsson:1994kr}, the $D$-dimensional ILST null string \cite{Isberg:1993av}, the $D$-dimensional Carroll-Weyl gauged string \cite{Sheikh-Jabbari:2026vqh, Sheikh-Jabbari:2026tpf}, and the $D$-dimensional hybrid null string \cite{Chen:2026klv}. By expressing all fields and constraint generators strictly in terms of mode expansions and adopting a unified algebraic framework, we analyze their quantum anomalies under two distinct worldsheet vacua: the induced vacuum and the flipped vacuum. With the BRST-compatible vacuum definition and the symmetric $\alpha=0$ zeta prescription, we show that no critical dimension is inferred from the vanishing of the quantum anomaly in the induced vacuum. In contrast,  the flipped highest-weight vacuum leads to non-trivial constraints,  reproducing the critical dimension $D=26$ for the ILST null strings,  a $\lambda$-dependent critical dimension $D(\lambda)$ for the hybrid null string whose range covers every positive integer $D\geq 4$ (reproduces $D=26$ at $\lambda=1$), and more importantly showing that the conformal string and the Carroll-Weyl gauged string are structurally anomalous with no consistent critical dimension due to the discrepancy of their central charge parameters $\tilde d_i$. Furthermore, the ILST null string model has target-space conformal symmetry $SO(D,2)$, the ghost-completed $SO(D,2)$ charges are closed on the induced-vacuum BRST cohomology in the $\alpha=0$ prescription, whereas the symmetry is quantum mechanically broken in the flipped vacuum.
}

\begin{document}
\raggedbottom
\maketitle

\section{Introduction}
\label{sec:intro}

The tensionless ($T = 0$) limit of string theory, often referred to as the null string, has attracted sustained interest as a potential window into the high-energy limit of string theory. An early hint came from the observation that, in the ultra-high-energy, fixed-angle regime, the string path integral is dominated by a saddle point with an enlarged symmetry~\cite{Gross:1987ar, Gross:1988ue}. Shortly after, it was shown that at the Hagedorn temperature $T_H$ the density of string states grows exponentially, signaling a phase transition~\cite{Atick:1988si}. More broadly, the tensionless limit connects to higher-spin gauge symmetries in string theory~\cite{Sundborg:2000wp}. It has subsequently been argued that in these high energy phases the effective string tension vanishes and the worldsheet Virasoro algebra undergoes an Inönü--Wigner contraction to the Bondi--Metzner--Sachs ($\mathfrak{bms}_3$) algebra~\cite{Barnich:2012aw, Bagchi:2012cy, Bagchi:2015nca, Bagchi:2021ban}.

The study of null strings has a long history, beginning with the classical formulation~\cite{Schild:1977vl} and a covariant action~\cite{Karlhede:1986pq}. The first comprehensive quantum treatment~\cite{Isberg:1993av}, hereafter referred to as the ILST null string, showed that the worldsheet Weyl invariance of the tensile string is replaced by a global target-space conformal symmetry $SO(D,2)$ in the tensionless limit. The light-cone quantization revealed that the physical spectrum collapses to diffeomorphism singlets, suggesting a topological phase with no local propagating degrees of freedom~\cite{Sundborg:1994aa}. To manifest the spacetime conformal symmetry on the worldsheet, the $D$-dimensional theory was embedded into a $(D+2)$-dimensional pseudo-Euclidean space and quantized via a Hamiltonian BRST procedure~\cite{Gustafsson:1994kr}, so it is referred to as the conformal string.

The past few years have witnessed a vigorous revival of this subject, driven by advances in Carrollian field theory and flat-space holography. The worldsheet metric of a null string is degenerate, so the worldsheet geometry is intrinsically Carrollian; indeed the $\mathfrak{bms}_3$ algebra is precisely the conformal Carroll algebra in two dimensions~\cite{Bagchi:2013bga}. This places tensionless strings within the broader framework of Carrollian approaches to flat-space holography~\cite{Bagchi:2012yk, Barnich:2010eb, Strominger:2013jfa, Donnay:2022aba}, where Carrollian structures on null infinity provide a natural arena for encoding gravitational scattering~\cite{Ciambelli:2019lap}. The tensionless limit was re-examined from the perspective of worldsheet Carrollian geometry~\cite{Bagchi:2020zmr, Bagchi:2021pci}, demonstrating that the choice of vacuum state---induced, flipped, or oscillator---radically alters the quantum consistency conditions~\cite{Bagchi:2021pci}. An overlooked local Carroll-Weyl gauge symmetry of the null string was recently identified~\cite{Sheikh-Jabbari:2026cnj, Sheikh-Jabbari:2026vqh, Sheikh-Jabbari:2026tpf}, which further refines the constraint algebra and modifies the ghost sector. Very recently, the geometric origin of this symmetry was clarified~\cite{Lindstrom:2026zno} by showing that the $D$-dimensional Carroll-Weyl gauged string arises as a Dirac reduction of the $(D+2)$-dimensional conformal string on a hypersurface. In the meanwhile, we~\cite{Chen:2026klv} introduced the hybrid null string, whose worldsheet symmetry is governed by a one-parameter family of algebras $\mathfrak{g}^{(1)}_\lambda$ that deforms $\mathfrak{bms}_3$, recovering it at $\lambda = 1$. A path-integral quantization of the Carroll-Weyl gauged string was performed~\cite{Duary:2026rlo}, confirming the need for an extended ghost system. Two recent short reviews~\cite{Lindstrom:2026quz, Sheikh-Jabbari:2026cnj} provided complementary perspectives on these developments, but a systematic, algebraic comparison of different formulations has been lacking.

In this paper, we conduct a unified BRST analysis of four distinct formulations of tensionless bosonic strings:
\begin{itemize}
  \item The $(D+2)$-dimensional conformal string~\cite{Gustafsson:1994kr}, with $\widehat{\mathfrak{sl}}(2,\mathbb{R}) \rtimes \textit{Vir}$ constraint algebra (hereafter $\textit{Vir}$ denotes the Virasoro algebra);
  \item The $D$-dimensional ILST null string~\cite{Isberg:1993av}, with $\mathfrak{bms}_3$ constraint algebra;
  \item The $D$-dimensional Carroll-Weyl gauged string~\cite{Sheikh-Jabbari:2026vqh, Sheikh-Jabbari:2026tpf}, with Carroll-Weyl constraint algebra;
  \item The $D$-dimensional hybrid null string~\cite{Chen:2026klv}, with $\mathfrak{g}^{(1)}_\lambda$ constraint algebra.
\end{itemize}
All four models share the same flat-gauge action $S = \frac{1}{2\pi} \int d^2\sigma \, \frac12 (\partial_\tau X)^2$, so the matter constraint generators $M^i_m$ of these models share the same form when they are defined. Their differences reside entirely in the ghost sectors, which encode the distinct residual gauge symmetries arising from the various gauge-fixing procedures. By expressing all fields and constraint generators strictly in terms of Fourier mode expansions and adopting a common Hamiltonian BRST formalism, we obtain a transparent, comparative picture of their quantum algebraic structures.

The worldsheet vacuum prescription plays a crucial role in our analysis. While Ref.~\cite{Bagchi:2021pci} classifies three vacua---induced, flipped, and oscillator---the oscillator vacuum imposes the physical state conditions in a highly non-standard manner and lies outside the scope of this work, we therefore focus on the two vacua with qualitatively distinct normal-ordering prescriptions:
\begin{enumerate}
  \item The \textbf{induced vacuum} $|0\rangle_{\text{Ind}}$, in which $p^M_n|0\rangle_{\text{Ind}}=0$ for every $n\in\mathbb Z$. The ghost field annihilation condition is model-dependent because we have to make sure that $Q_B|0\rangle_{\text{Ind}}=0$. The conformal string has the BRST-compatible vacuum $b_{-1,n}|0\rangle_{\text{Ind}}=c^1_n|0\rangle_{\text{Ind}}=0$, while the other three strings use an antighost-annihilation vacuum, $b_{i,n}|0\rangle_{\text{Ind}}=0$.
  \item The \textbf{flipped vacuum} $|0\rangle_{\text{Flip}}$, a highest-weight vacuum where only positive-frequency modes annihilate. Here, non-vanishing two-point functions produce finite double contractions that contribute genuine quantum central extensions to the constraint algebra, requiring the cancellation of anomalies for a consistent quantum theory.
\end{enumerate}
The induced-vacuum calculations contain divergent sums. We use the symmetric $\alpha=0$ member of the generalized zeta family of Ref.~\cite{Hwang:1998brst}. The reason why we adopt the special choice of $\alpha=0$ is that the quantum anomalies with general $D$ and non-zero $\alpha$ do not satisfy the Jacobi identity.

The paper is organized as follows. Section~\ref{sec:four_formulations} presents the four formulations in a unified notation, detailing their gauge symmetries, gauge-fixing procedures, residual constraint algebras, and the Fourier-mode expansions of all matter constraint generators. Section~\ref{sec:BRST} constructs the BRST charge, ghost constraint generators, and the full quantum constraint algebra for each model, emphasizing the role of Jacobi identities in constraining the central extensions. Section~\ref{sec:anomalies} analyzes quantum anomalies under the induced and flipped vacua, deriving the critical dimensions and anomaly structures summarized in Table~\ref{tab:critical_dims}. Section~\ref{sec:conclusion} contains our conclusions and outlook.

The Appendix~\ref{app:wick} reviews the Wick theorem formalism and the double-contraction mechanism that underlie all quantum anomaly computations. The Appendices ~\ref{app:induced} and ~\ref{app:flipped_derivations} provide the exhaustive derivations of all central extension coefficients under the induced and flipped vacua, respectively. The Appendix~\ref{app:SO(D,2)} is a summary of the ILST null string model. It shows that the complete ghost-extended algebra $SO(D,2)$ is closed in the induced-vacuum BRST cohomology in the $\alpha=0$ prescription, while the fixed special-conformal representative is anomalous in the flipped vacuum even at $D=26$. This provides an explicit realization of a general mechanism: in theories with a degenerate worldsheet metric, anomalies can be pushed from the gauge sector into the global symmetry sector without rendering the BRST charge itself non-nilpotent.

\noindent\textbf{Note:} While we were finishing the project, we learned  that  Sarthak Duary and Sourav Maji \cite{Duary:2026cw} were working independently on the Carroll-Weyl gauged string and successively performed BRST quantization through the Faddeev-Popov ghost procedure. Obviously their work has substantial overlap with our study on Carroll-Weyl gauged string.

\section{Four formulations of tensionless strings in unified notation}
\label{sec:four_formulations}
To formulate tensionless string theories, we consider the closed string worldsheet coordinates $(\tau, \sigma)$ with $\sigma \in [0, 2\pi]$. Throughout all four models, \(\alpha,\beta\in\{\tau,\sigma\}\) denote worldsheet indices, while \(m,n,k,l\in\mathbb Z\) label Fourier modes; a prime denotes \(\partial_\sigma\), and the sums of modes without displayed limits run over \(\mathbb Z\). As will be shown, the action of the four different models takes the same form:
\begin{equation}\label{eq:flat_gauge_action}
    S_{\text{flat gauge}}=\frac{1}{2\pi}\int d^2\sigma \frac{1}{2}(\p_\tau X)^2
\end{equation}
after imposing some gauge choices. By solving the equation of motion $\partial_\tau^2 X^M = 0$, the target-space coordinates $X^M(\tau, \sigma)$ and their conjugate momenta $P^M(\tau, \sigma)$ are expanded in Fourier modes as
\begin{equation}\label{eq:fourier_expansion}
X^M(\tau, \sigma) = \sum_{m=-\infty}^{+\infty} x^M_m e^{-im\sigma} + \tau \sum_{n=-\infty}^{+\infty} p^M_n e^{-in\sigma}, \quad P^M(\tau, \sigma) = \frac{1}{2\pi} \sum_{n=-\infty}^{+\infty} p^M_n e^{-in\sigma}
\end{equation}
The equal-time canonical commutation relations
\begin{equation}
[X^M(\tau, \sigma), P^N(\tau, \sigma')] = i \delta(\sigma - \sigma') \eta^{MN}
\end{equation}
translate to the following mode commutation relations:
\begin{equation}\label{eq:canonical_commutator}
[x^M_m, p^N_n] = i\delta_{m+n, 0} \eta^{MN}
\end{equation}
where $M, N$ run over the target-space coordinates, and $\eta^{MN}$ is the target-space metric.

In the canonical Hamiltonian formalism, the physical states are restricted by first class constraints. For the classical ILST null string, the constraints are $P^2 = 0$ and $P \cdot X' = 0$, which are generated by the modes $M^{-1}_m$ and $M^L_m$, respectively. In the conformal string, the constraints are augmented by $X^2 = 0$ and $X \cdot P = 0$, generated by $M^1_m$ and $M^0_m$. The general mode expressions for these constraints $M^i_m$ ($i \in \{-1, 0, 1, L\}$) are given in the following subsections.

These constraints can be read out from the first-order actions before gauge fixing. The conformal string action can be written in first-order form as
\begin{equation*}
S_{\mathrm c}=\int d\tau d\sigma\,
\left[P\cdot\dot X-eM^{-1}+rM^L+2AM^0+\Phi M^1\right],
\end{equation*}
where $e=(V^\tau)^{-2}$, $r=V^\sigma/V^\tau$ and $A=W_\tau+rW_\sigma$, with $V^\tau\ne0$. In $D$ target dimensions, the ILST first-order string action follows from this expression by omitting the $A$ and $\Phi$ terms, while the Carroll--Weyl gauged string action follows by omitting only the $\Phi$ term. The momenta conjugate to the Lagrange multipliers vanish, and preservation of these primary constraints gives the corresponding first-class secondary constraints. Independent variation of $A$ gives the scaling Gauss-law constraint $M^0$. The remaining connection component has a primary constraint generating its shift symmetry. The hybrid action in first-order form and its constraints will be discussed separately below.

These constraints also appear in the charges associated with residual gauge symmetries. On the closed circle, they are linear combinations of these first-class constraints and hence vanish weakly, with no boundary contribution. The corresponding time-dependent conserved charges are $Q(\tau;X,P)=Q(0;X-2\pi\tau P,P)$, satisfying $\partial_\tau Q+\{Q,H_{\rm flat}\}_{\rm PB}=0$, where $H_{\rm flat}=\int d\sigma\,\pi P^2$. Below, we present the gauge-fixing procedure for the conformal string and the corresponding results for the other three models.

\subsection{The \texorpdfstring{$(D+2)$-dimensional}{(D+2)-dimensional} conformal string}
The conformal string \cite{Gustafsson:1994kr} manifests target-spacetime conformal symmetry by embedding the $D$-dimensional physical target space into a $(D+2)$-dimensional pseudo-Euclidean space with signature $(-+++ \dots + -)$ where $M, N = 0, \dots, D+1$. The action of the conformal string is of the form
    \begin{equation}
    S_{\mathrm c} = \frac{1}{2\pi} \int d^2\sigma \frac{1}{2} \left( V^\alpha V^\beta D_\alpha X^M D_\beta X^N \eta_{MN} + \Phi X^2 \right)
    \end{equation}
    where $V^\alpha$ is a vector density of weight $1/2$, $\Phi$ is a Lagrange multiplier enforcing $X^2 = 0$, and scale-covariant derivatives are $D_\alpha = \partial_\alpha + W_\alpha$ with scale connection $W_\alpha$.
    It has been shown in \cite{Gustafsson:1994kr} that the $(D+2)$-dimensional conformal string is classically equivalent to the $D$-dimensional ILST null string by integrating out $W_\alpha$ and $\Phi$.

    The action $S_{\mathrm c}$ is invariant under the following local gauge transformations:
    \begin{itemize}
        \item \textbf{Worldsheet diffeomorphisms} parameterized by $\epsilon^\alpha$:
        \begin{align}
        \delta_\epsilon X^M &= \epsilon^\alpha \partial_\alpha X^M, \\
        \delta_\epsilon V^\alpha &= \epsilon^\beta \partial_\beta V^\alpha - V^\beta \partial_\beta \epsilon^\alpha + \frac{1}{2} (\partial_\beta \epsilon^\beta) V^\alpha, \\
        \delta_\epsilon W_\alpha &= \epsilon^\beta \partial_\beta W_\alpha + W_\beta \partial_\alpha \epsilon^\beta, \\
        \delta_\epsilon \Phi &= \epsilon^\alpha \partial_\alpha \Phi + (\partial_\alpha \epsilon^\alpha) \Phi.
        \end{align}
        \item \textbf{Local scale transformations} parameterized by $\Lambda$:
        \begin{equation}
        \delta_\Lambda X^M = -\Lambda X^M, \quad \delta_\Lambda W_\alpha = \partial_\alpha \Lambda, \quad \delta_\Lambda V^\alpha = \Lambda V^\alpha, \quad \delta_\Lambda \Phi = 2\Lambda \Phi.
        \end{equation}
        \item \textbf{Connection shift symmetry} parameterized by a vector field $f_\alpha$ orthogonal to $V^\alpha$ ($V^\alpha f_\alpha = 0$):
        \begin{equation}
        \delta_f W_\alpha = f_\alpha.
        \end{equation}
        \item \textbf{Additional gauge transformation associated with the gauge field $\Phi$} parameterized by $\theta$:
        \begin{equation}
        \label{eq:gauge_transform_X2}
        \delta_\theta X^M = 0, \quad \delta_\theta W_\alpha = -\frac{1}{2} U_\alpha \theta, \quad \delta_\theta \Phi = -\frac{1}{2} \partial_\alpha (V^\alpha \theta) + V^\alpha W_\alpha \theta
        \end{equation}
        where $U_\alpha$ is an auxiliary vector field satisfying $V^\alpha U_\alpha = 1$.
    \end{itemize}
    We can now discuss how to perform complete gauge-fixing using the local gauge symmetries, identify the residual symmetries, and establish their correspondence with the matter constraints. The action $S_{\mathrm c}$ possesses five local gauge parameters: two from worldsheet diffeomorphisms, one from local scale transformations, one from the connection-shift parameter $f_\alpha$, subject to $V^\alpha f_\alpha = 0$, and one from the additional gauge symmetry associated with the gauge field $\Phi$. Correspondingly, there are five gauge field components to be fixed: two from the vector density $V^\alpha$, two from the scale connection $W_\alpha$, and one from the Lagrange multiplier $\Phi$. The matching of the number of gauge parameters and gauge field components allows us to completely fix all gauge fields to the flat gauge:
    \begin{equation}
    V^\alpha = (1, 0)^T, \quad W_\alpha = 0, \quad \Phi = 0.
    \end{equation}
    Specifically, the gauge fixing is achieved by considering the combined effect of all local gauge transformations. Under the combined local gauge transformations parameterized by $\epsilon^\alpha$, $\Lambda$, $f_\alpha$ (with $V^\alpha f_\alpha = 0$), and the additional gauge parameter $\theta$, the total variations of the gauge fields are:
    \begin{align}
    \delta V^\alpha &= \epsilon^\beta \partial_\beta V^\alpha - V^\beta \partial_\beta \epsilon^\alpha + \frac{1}{2} (\partial_\beta \epsilon^\beta) V^\alpha + \Lambda V^\alpha \\
    \delta W_\alpha &= \epsilon^\beta \partial_\beta W_\alpha + W_\beta \partial_\alpha \epsilon^\beta + \partial_\alpha \Lambda + f_\alpha - \frac{1}{2} U_\alpha \theta \\
    \delta \Phi &= \epsilon^\alpha \partial_\alpha \Phi + (\partial_\alpha \epsilon^\alpha) \Phi + 2\Lambda \Phi - \frac{1}{2} \partial_\alpha (V^\alpha \theta) + V^\alpha W_\alpha \theta.
    \end{align}
    
    The residual worldsheet gauge symmetries are defined by the subgroup of local transformations that preserve the flat gauge, meaning that the total variations of the gauge fields must vanish simultaneously:
    \begin{equation}
    \delta V^\alpha = 0, \quad \delta W_\alpha = 0, \quad \delta \Phi = 0.
    \end{equation}
    Evaluating these combined variations under the gauge-fixed background $V^\alpha = (1, 0)^T$, $W_\alpha = 0$, and $\Phi = 0$, we find:
    \begin{itemize}
        \item For $V^\alpha$, since $V^\tau = 1$ and $V^\sigma = 0$, we have:
        \begin{align}
        \delta V^\sigma &= - \partial_\tau \epsilon^\sigma = 0 \implies \epsilon^\sigma = \xi(\sigma), \\
        \delta V^\tau &= - \partial_\tau \epsilon^\tau + \frac{1}{2} (\partial_\tau \epsilon^\tau + \partial_\sigma \epsilon^\sigma) + \Lambda = 0 \implies \Lambda = \frac{1}{2}\partial_\tau \epsilon^\tau - \frac{1}{2}\partial_\sigma \epsilon^\sigma,
        \end{align}
        which shows that the spatial diffeomorphism parameter $\epsilon^\sigma$ depends only on $\sigma$, and the scale parameter $\Lambda$ is determined by the diffeomorphism parameter $\epsilon^\tau$.
        \item For $W_\alpha$, the orthogonality condition $V^\alpha f_\alpha = 0$ restricts the shift parameter to $f_\tau = 0$. Choosing the auxiliary vector field $U_\alpha = (1, 0)$ so that $U_\tau = 1$ and $U_\sigma = 0$, we obtain:
        \begin{align}
        \delta W_\tau &= \partial_\tau \Lambda - \frac{1}{2} \theta = 0 \implies \theta = 2\partial_\tau \Lambda, \\
        \delta W_\sigma &= \partial_\sigma \Lambda + f_\sigma = 0 \implies f_\sigma = -\partial_\sigma \Lambda.
        \end{align}
        This shows that the additional parameter $\theta$ and the connection shift parameter $f_\sigma$ are completely determined by $\Lambda$.
        \item For $\Phi$, the variation reduces to:
        \begin{equation}
        \delta \Phi = - \frac{1}{2} \partial_\tau \theta = 0 \implies \partial_\tau \theta = 0.
        \end{equation}
    \end{itemize}
    Combining these requirements, the condition $\partial_\tau \theta = 0$ implies $\partial_\tau^2 \Lambda = 0$. Since $\Lambda = \frac{1}{2}\partial_\tau \epsilon^\tau - \frac{1}{2}\partial_\sigma \epsilon^\sigma$ and $\epsilon^\sigma = \xi(\sigma)$, this restricts the diffeomorphism parameter to satisfy $\partial_\tau^3 \epsilon^\tau = 0$, yielding:
    \begin{equation}
    \epsilon^\tau(\tau, \sigma) = \eta(\sigma) + \tau \rho(\sigma) + \tau^2 \omega(\sigma).
    \end{equation}
    Consequently, the residual worldsheet gauge symmetries are parameterized by four independent functions of $\sigma$: $\eta(\sigma)$, $\xi(\sigma)$, $\rho(\sigma)$, and $\omega(\sigma)$. On the physical state space defined on the matter fields $(X^M, P^M)$, the generators of these residual gauge transformations are given by the conserved Noether charges. Under the combined residual transformations, the variations of the matter fields are generated by the Noether charge:
    \begin{equation}
    Q_{\text{Noether}} = \int_0^{2\pi} d\sigma \left[ \eta(\sigma) M^{-1}(\sigma) - \rho(\sigma) M^0(\sigma) + \omega(\sigma) M^1(\sigma) + \xi(\sigma) \left( M^L(\sigma) - \partial_\sigma M^0(\sigma) \right) \right].
    \end{equation}
    In the quantum theory, the residual transformations of the matter fields are generated by the Noether charge via the Heisenberg commutator: $\delta X^M = -i [ X^M, Q_{\text{Noether}} ]$. Here, we define the continuous matter constraint fields $M^i(\sigma)$ as:
    \begin{align}
    M^{-1}(\sigma) \equiv \pi P^2(\sigma), ~~~~
    M^0(\sigma) \equiv \frac{1}{2} X(\sigma) \cdot P(\sigma), \nonumber\\
    M^1(\sigma) \equiv \frac{1}{4\pi} X^2(\sigma), ~~~~
    M^L(\sigma) \equiv P(\sigma) \cdot X'(\sigma).\nonumber
    \end{align}
    By defining the Fourier modes of these constraints as $M^i_m = \int_0^{2\pi} d\sigma e^{im\sigma} M^i(\sigma)$, we can express them in terms of the modes of the matter fields $X$ and $P$ as:
    \begin{align}
    M^{-1}_m = \frac{1}{2} \sum_{k=-\infty}^{+\infty} \mathopen{:} p_k \cdot p_{m-k} \mathclose{:},~~~~
    M^0_m = \frac{1}{2} \sum_{k=-\infty}^{+\infty} \mathopen{:} x_k \cdot p_{m-k} \mathclose{:}, \nonumber\\
    M^1_m = \frac{1}{2} \sum_{k=-\infty}^{+\infty} \mathopen{:} x_k \cdot x_{m-k} \mathclose{:},~~~~ 
    M^L_m = -i \sum_{k=-\infty}^{+\infty} k \mathopen{:} x_k \cdot p_{m-k} \mathclose{:}.\nonumber
    \end{align}
    The matter constraint generators satisfy the $\widehat{\mathfrak{sl}}(2, \mathbb{R}) \rtimes \textit{Vir}$ algebra:
    \begin{align}
    [M^1_m, M^{-1}_n] &= 2i M^0_{m+n}, \quad
    [M^L_m, M^L_n] = (m-n) M^L_{m+n}, \quad
    [M^0_m, M^L_n] = m M^0_{m+n}, \nonumber\\
    [M^1_m, M^0_n] &= i M^1_{m+n}, \quad
    [M^{-1}_m, M^0_n] = -i M^{-1}_{m+n}, \nonumber\\
    [M^1_m, M^L_n] &= (m+n) M^1_{m+n}, \quad
    [M^{-1}_m, M^L_n] = (m-n) M^{-1}_{m+n}.
    \end{align}
    These are only classical relations, and their quantum versions, which may contain central extensions due to  normal ordering, will be analyzed in Sections~\ref{sec:BRST} and~\ref{sec:anomalies}.

\subsection{The \texorpdfstring{$D$-dimensional}{D-dimensional} ILST null string}
The ILST null string~\cite{Isberg:1993av} is obtained by directly taking $T \to 0$ in the Nambu-Goto action.  The action of the ILST null string is given by:
    \begin{equation}
    S_{\mathrm n} = \frac{1}{2\pi} \int d^2\sigma \frac{1}{2} V^{\alpha} V^{\beta} \partial_{\alpha} X^\mu \partial_{\beta} X^\nu \eta_{\mu\nu},\hspace{3ex}\mu, \nu = 0, \dots, D-1, 
    \end{equation}
    where \(V^\alpha\) is a vector density of weight $1/2$.

 For the ILST null string the only local symmetry is worldsheet diffeomorphism invariance.  Fixing \(V^\alpha=(1,0)^T\) and imposing \(\delta V^\alpha=0\) gives directly
\begin{equation}
\epsilon^\sigma=\xi(\sigma),
\qquad
\epsilon^\tau=\eta(\sigma)+\tau\xi'(\sigma).
\end{equation}
The two residual functions multiply the translation and Virasoro constraints,
\begin{equation}
Q_{\rm Noether}=\int_0^{2\pi}d\sigma\,
\bigl(\eta M^{-1}+\xi M^L\bigr).
\end{equation}
Thus the ILST null string retains the conformal string mode operators $M^{-1}_m$ and $M^L_m$ (with target indices restricted to $\mu=0,\ldots,D-1$), and their non-vanishing brackets are
\begin{equation}
[M^L_m,M^L_n]=(m-n)M^L_{m+n},
\qquad
[M^L_m,M^{-1}_n]=(m-n)M^{-1}_{m+n}.
\end{equation}
Together with $[M^{-1}_m,M^{-1}_n]=0$, they give rise to the $\mathfrak{bms}_3$ algebra.

\subsection{The \texorpdfstring{$D$-dimensional}{D-dimensional} Carroll--Weyl gauged string}
The Carroll--Weyl gauged string gauges local scale transformations with a connection \(W_\alpha\)~\cite{Sheikh-Jabbari:2026vqh, Sheikh-Jabbari:2026tpf}:
\begin{equation}
S_{\mathrm g}=\frac{1}{2\pi}\int d^2\sigma\,
\frac12 V^{\alpha}V^{\beta}D_{\alpha}X\cdot D_{\beta}X,
\qquad D_{\alpha}X^\mu=\partial_{\alpha}X^\mu+W_{\alpha}X^\mu.
\end{equation}
The worldsheet diffeomorphisms generated by $\epsilon^\alpha$, local scale transformations generated by $\Lambda$, and the connection shift \(\delta_fW_\alpha=f_\alpha\) with \(V^\alpha f_\alpha=0\) fix \(V^\alpha=(1,0)^T\) and \(W_\alpha=0\).  The transformations preserving this gauge are summarized below
\begin{equation}
\epsilon^\sigma=\xi(\sigma),\qquad
\epsilon^\tau=\eta(\sigma)+\tau\bigl(\xi'(\sigma)+\rho(\sigma)\bigr),\qquad
\Lambda=\frac{\rho}{2},\qquad
f_\tau=0,\quad f_\sigma=-\frac{\rho'}{2}.
\end{equation}
Accordingly, there is
\begin{equation}
Q_{\rm Noether}=\int_0^{2\pi}d\sigma\,
\bigl(\eta M^{-1}-\rho M^0+\xi M^L\bigr).
\end{equation}
The three mode operators are the corresponding conformal-string operators restricted to $D$ target dimensions.  The generators of the algebra have the commutation relations 
\begin{align}
[M^L_m,M^L_n]&=(m-n)M^L_{m+n},
& [M^0_m,M^L_n]&=mM^0_{m+n},\nonumber\\
[M^{-1}_m,M^0_n]&=-iM^{-1}_{m+n},
& [M^{-1}_m,M^L_n]&=(m-n)M^{-1}_{m+n},
\end{align}
with $[M^0_m,M^0_n]=[M^{-1}_m,M^{-1}_n]=0$.  This generates the Carroll-Weyl algebra.

\subsection{The \texorpdfstring{$D$-dimensional}{D-dimensional} hybrid null string}
The hybrid model~\cite{Chen:2026klv} admits a first-order phase-space formulation for every $\lambda\in\mathbb R$. For $\lambda\ne-1$, it is locally equivalent to the covariant second-order action
\begin{equation}
S_{\mathrm h}=\frac{1}{4\pi}\int d^2\sigma\,\mathcal K\cdot\mathcal K,
\qquad
\mathcal K^\mu=V^\alpha\partial_\alpha X^\mu
+a(\partial_\alpha V^\alpha)X^\mu.
\label{eq:S_Stueckelberg}
\end{equation}
Here $X^\mu$ and $V^\alpha$ have density weights $\Delta/2$ and
$(1-\Delta)/2$, respectively, making $\mathcal K$ a scalar density of
weight $1/2$. This covariant parametrization uses
\begin{equation}
\lambda=\frac{1-\Delta}{1+\Delta},\qquad
 a=\frac{\Delta}{1+\Delta}=\frac{1-\lambda}{2},\qquad \Delta\in\mathbb R\setminus\{-1\}.
\label{eq:lambda_Delta}
\end{equation}
To obtain the phase-space action, set $u=V^\tau>0$, $Y=u^aX$,
$r=V^\sigma/u$, and $e=u^{-2(1-a)}=u^{-(1+\lambda)}$.
Then $\mathcal K=u^{1-a}(\dot Y+rY'+ar'Y)$ and
$\Pi=(\dot Y+rY'+ar'Y)/(2\pi e)$. The Legendre transformation and an
integration by parts on the closed circle give
\begin{equation*}
\begin{aligned}
S_{\mathrm h}^{(1)}&=\int d\tau d\sigma\,
\left[\Pi\cdot\dot Y-e\mathcal H+r\mathcal D_\lambda\right],\\
\mathcal H&=\pi\Pi^2,\qquad
\mathcal D_\lambda=\Pi\cdot Y'-\frac{1-\lambda}{2}(\Pi\cdot Y)'.
\end{aligned}
\end{equation*}
We take $e$ and $r$ to be independent Lagrange multipliers in
$S_{\mathrm h}^{(1)}$, which is well defined also at $\lambda=-1$.
Their conjugate momenta obey the primary constraints $p_e\approx0$ and
$p_r\approx0$. With the total Hamiltonian
$H_T=\int d\sigma\,(e\mathcal H-r\mathcal D_\lambda+\nu_ep_e+\nu_rp_r)$,
preservation of the primary constraints gives
\begin{equation*}
\dot p_e=-\mathcal H\approx0,\qquad
\dot p_r=\mathcal D_\lambda\approx0.
\end{equation*}
Their Poisson brackets close for every $\lambda$, so these secondary constraints
are first class, with no independent scaling constraint. Their Fourier modes give
the classical counterpart of~\eqref{eq:model4_alg}.
At $\lambda=-1$, the relation $e=u^{-(1+\lambda)}$ becomes $e=1$ and is
not invertible. Thus $S_{\mathrm h}^{(1)}$ with independent $e$ defines this case;
simply setting $a=1$ in the second-order expression does not impose the
zero mode of $\mathcal H$. In flat gauge, $e=1$ and $r=0$,
we identify $Y=X$ and $\Pi=P$, so that
$\mathcal H=M^{-1}$ and $\mathcal D_\lambda=M^{L_\lambda}$.
Target-space translations are generally absent for $a\ne0$
($\delta\mathcal K=a(\partial_\alpha V^\alpha)b$ under $\delta X=b$),
so anomaly cancellation in the model defined by~\eqref{eq:S_Stueckelberg} alone does not establish the
physical spectrum of a Poincar\'e-invariant string theory.
In this gauge, the residual transformations and charges are\begin{equation}
\epsilon^\sigma=\xi(\sigma),\qquad
\epsilon^\tau=\eta(\sigma)+\lambda\tau\xi'(\sigma),\qquad
Q_{\rm Noether}=\int_0^{2\pi}d\sigma\,
\bigl(\eta M^{-1}+\xi M^{L_\lambda}\bigr).
\end{equation}
The only new matter generator is
\begin{align}
M^{L_\lambda}(\sigma)
&=M^L(\sigma)+(\lambda-1)\partial_\sigma M^0(\sigma).\nonumber\end{align}
Here \(M^0=\tfrac12X\cdot P\) is used only as the kinematic bilinear entering \(M^{L_\lambda}\) and is not an additional hybrid-string constraint. This new generator can be expanded as
\begin{align}
M^{L_\lambda}_m
&=-i\sum_{k=-\infty}^{+\infty}
\left[k+m\frac{\lambda-1}{2}\right]
\mathopen{:}x_k\cdot p_{m-k}\mathclose{:}.
\end{align}
The expansion of $M^{-1}_m$ is unchanged. The modes in $M^{L_\lambda}$ and $M^{-1}$ obey the commutation relations
\begin{equation}
[M^{L_\lambda}_m,M^{L_\lambda}_n]=(m-n)M^{L_\lambda}_{m+n},\qquad
[M^{L_\lambda}_m,M^{-1}_n]=(\lambda m-n)M^{-1}_{m+n},\qquad
[M^{-1}_m,M^{-1}_n]=0,
\label{eq:model4_alg}
\end{equation}
which defines the algebra $\mathfrak g^{(1)}_\lambda$.  

As we showed above, even though the four models share the same flat-gauge matter action, the choice of residual constraints are different. Hence the ghost sectors, which will be analyzed shortly, distinguish them quantum mechanically.

\section{BRST quantization and quantum constraint algebra}
\label{sec:BRST}

In the Hamiltonian BRST formalism, one promotes each first-class constraint $M^i_m$ to a quantum operator and introduces a pair of canonically conjugate Grassmann-odd ghosts: a ghost $c^i_m$ (ghost number $+1$) and an antighost $b_{i,m}$ (ghost number $-1$), satisfying
\begin{equation}\label{eq:ghost_anticommutator}
\{b_{i, m}, c^j_n\} = \delta_i^j \delta_{m+n, 0}.
\end{equation}
The index $i$ labels the constraint sectors: $i \in \{-1, 0, 1, L\}$ for the conformal string, $i \in \{-1, L\}$ for the ILST null string, $i \in \{-1, 0, L\}$ for the Carroll--Weyl gauged string, and $i \in \{-1, L_\lambda\}$ for the hybrid null string.
Let \(\mathcal A=(i,m)\) be a composite constraint label and denote
\begin{equation}
M_{(i,m)}=M^i_m,\qquad
c^{(i,m)}=c^i_{-m},\qquad
b_{(i,m)}=b_{i,m}.
\end{equation}
Then we can rewrite the matter constraint algebra as
\([M_{\mathcal A},M_{\mathcal B}]=f_{\mathcal A\mathcal B}{}^{\mathcal C}M_{\mathcal C}\).
The summation rule is that repeated composite labels include both the sum over sectors  and the sum over modes \(m\in\mathbb Z\).  

At the classical level, before choosing a vacuum and a normal-ordering prescription, the closure of the constraint algebra fixes the following minimal Grassmann-odd BRST charge of ghost number \(+1\):
\begin{equation}\label{eq:Q0_general}
Q_0=c^{\mathcal A} M_{\mathcal A}
-\frac{1}{2}f_{\mathcal A\mathcal B}{}^{\mathcal C}
c^{\mathcal A} c^{\mathcal B} b_{\mathcal C}.
\end{equation}
The coefficient of the second term is uniquely determined by $Q_0^2=0$. This construction assumes a closed, irreducible constraint algebra and the minimal cubic ansatz, which is correct for these four models since the BRST invariance with $\delta_B^2=0$ holds off-shell. The total constraint operator is defined as 
\begin{equation}
    H^i_m \equiv \{b_{i,m},Q_0\}=M^i_m+G^i_m.
\end{equation}

At the quantum level, one needs to take into account of the normal ordering, which induce constants to the zero modes of the quantum constraints.  Once \(H^i_0\) is shifted to \(H^i_0-A_i\), where $A_i$'s are normal-ordering zero-mode constants, the BRST charge must acquire the corresponding term \(-A_i c^i_0\),  
\begin{align}
Q &\equiv Q_0-\sum_i A_i c^i_0, 
\end{align}
and the total constraint operator can still be defined by
\begin{align}
\tilde\phi^i_m &\equiv \{b_{i,m},Q\}=H^i_m-A_i\delta_{m,0}.
\end{align}
  Here $\tilde{\phi}^i_m$ is the total constraint operator at the quantum level. Throughout this paper, we refer to the individual constraint generators $\tilde{\phi}^i_m$ by their physical roles in the underlying algebra: $\tilde{\phi}^{-1}_m$ as the \textbf{worldsheet supertranslation constraint}, $\tilde{\phi}^0_m$ as the \textbf{worldsheet scaling constraint}, $\tilde{\phi}^1_m$ as the \textbf{worldsheet special-conformal constraint}, and $\tilde{\phi}^L_m$ (or $\tilde{\phi}^{L_\lambda}_m$) as the \textbf{Virasoro constraint}. These names apply across all models where the corresponding generators are present.

While the matter constraints $M^i_m$ remain identical across the four formulations wherever they are defined, the Faddeev--Popov ghost sectors differ because of distinct gauge fixings. We use the descriptive subscripts $\{\mathrm c,\mathrm n,\mathrm g,\mathrm h\}$ for the conformal string, ILST null string, Carroll--Weyl gauged string, and hybrid null string, respectively. Thus their unshifted BRST charges are $Q_{\mathrm c,0}$, $Q_{\mathrm n,0}$, $Q_{\mathrm g,0}$, and $Q_{\mathrm h,0}$, with corresponding shifted charges $Q_{\mathrm c}$, $Q_{\mathrm n}$, $Q_{\mathrm g}$, and $Q_{\mathrm h}$; the same subscripts distinguish $G^i_m$ and $\tilde{\phi}^i_m$ when needed.
However, for each model, the labels on \(c^i_m,b_{i,m},A_i,d^{ij}_m\), and \(\tilde d_r\) are suppressed.

In mode notation, define \(U^{ij}{}_{r}(m,n)\) by the matter algebra without extension
\[
[M^i_m,M^j_n]=U^{ij}{}_{r}(m,n)M^r_{m+n}.
\]
The corresponding quantum algebra could be of the form
\begin{equation}\label{eq:quantum_algebra}
[\tilde{\phi}^i_{m}, \tilde{\phi}^j_{n}]
=U^{ij}{}_{r}(m, n) \tilde{\phi}^r_{m+n}
+d^{ij}_m \delta_{m+n,0}.
\end{equation}
The antisymmetry of the commutator implies \(d^{ij}_m=-d^{ji}_{-m}\). For the composite indices \(\mathcal A=(i,m)\), \(\mathcal B=(j,n)\), and \(\mathcal C=(r,p)\) used in Eq.~\eqref{eq:Q0_general}, the structure coefficients in two notations are related explicitly by
\begin{equation}
f_{(i,m)(j,n)}{}^{(r,p)}
=U^{ij}{}_{r}(m,n)\,\delta_{p,m+n}.
\label{eq:f_U_dictionary}
\end{equation}
Thus \(U^{ij}{}_{r}(m,n)\) contains the non-trivial mode-dependent coefficient, while the Kronecker delta in \(f_{\mathcal A\mathcal B}{}^{\mathcal C}\) enforces mode conservation. The central terms $d^{ij}_m$ are heavily restricted by the quantum Jacobi identities:
\begin{equation}
[[\tilde{\phi}^i_{m}, \tilde{\phi}^j_{n}], \tilde{\phi}^k_{l}] + [[\tilde{\phi}^k_{l}, \tilde{\phi}^i_{m}], \tilde{\phi}^j_{n}] + [[\tilde{\phi}^j_{n}, \tilde{\phi}^k_{l}], \tilde{\phi}^i_{m}] = 0.
\end{equation}

From a mathematical perspective, these central extensions are classified by the second Lie algebra cohomology group $H^2(\mathfrak{g}, \mathbb{C})$. A central extension is a trivial coboundary if it can be absorbed by shifting the zero-modes of the constraint generators,
\[
\tilde{\phi}^i_m \longmapsto \tilde{\phi}^{\prime i}_m
=\tilde{\phi}^i_m+s_i\delta_{m,0}.
\]
The shift parameter \(s_i\) is not independent of the normal-ordering constant \(A_i\).  If \(A_i\) is changed to \(A'_i=A_i+\delta A_i\), then
\[
\tilde\phi^{\prime i}_m=H^i_m-A'_i\delta_{m,0}
=\tilde\phi^i_m-\delta A_i\delta_{m,0},
\qquad s_i=-\delta A_i.
\]
In particular, relative to the unshifted operator \(H^i_m\), one has \(s_i=-A_i\).  Consequently the central terms transform consistently as
\[
d^{\prime ij}_m=d^{ij}_m-U^{ij}{}_{r}(m,-m)s_r
=d^{ij}_m+U^{ij}{}_{r}(m,-m)\delta A_r.
\]
The extensions that cannot be removed by such shifts correspond to non-trivial cohomology classes, representing genuine physical quantum anomalies.

In the presence of the quantum anomalies $d^{i j}_m$, the square of the BRST charge is
\begin{equation}\label{eq:Q2_anomaly}
Q^2=\frac{1}{2}\sum_{m\in\mathbb Z}\sum_{i,j}d^{ij}_m\,c^i_{-m}c^j_m.
\end{equation}
The sum includes $m=0$.  Quantum nilpotency requires all coefficients in this expression to vanish.  Trivial cocycles in $d^{i j}_m$ are cancelled by a choice of $A_i$, while non-trivial cohomology classes cannot be removed in this way and lead to the critical dimension and $\lambda$ conditions for the hybrid null string, as analyzed below.

From a group-theoretic perspective, the constraint algebras of the ILST null, Carroll--Weyl gauged, and hybrid null strings are all subalgebras of the $\widehat{\mathfrak{sl}}(2,\mathbb{R}) \rtimes \textit{Vir}$ algebra of the conformal string, obtained from the $\widehat{\mathfrak{sl}}(2)$ factor through successive truncations and deformation:
\[
\begin{array}{c}
\widehat{\mathfrak{sl}}(2,\mathbb{R}) \rtimes \textit{Vir} \quad \text{(conformal string, 4 generators)} \\[4pt]
\Big\downarrow \text{drop } M^1 \\[4pt]
\widehat{\mathfrak{aff}}(1) \rtimes \textit{Vir} \quad \text{(Carroll--Weyl gauged string, 3 generators)} \\[4pt]
\Big\downarrow \text{drop } M^0 \\[4pt]
\widehat{\mathfrak{u}}(1) \rtimes \textit{Vir} \cong \mathfrak{bms}_3 \quad \text{(ILST null string, 2 generators)} \\[4pt]
\Big\downarrow \text{deform } \rtimes \text{ to } \rtimes_\lambda \\[4pt]
\widehat{\mathfrak{u}}(1) \rtimes_\lambda \textit{Vir} \cong \mathfrak{g}^{(1)}_\lambda \quad \text{(hybrid null string, 2 generators)}
\end{array}
\]
Dropping $M^1$ (the special-conformal generator) truncates $\widehat{\mathfrak{sl}}(2)$ to its Borel subalgebra, the loop extension of the two-dimensional affine algebra of the line, \(\mathfrak{aff}(1)\), spanned by $\{M^{-1}, M^0\}$.  Dropping $M^0$ (the scaling generator) further reduces this to the Abelian worldsheet supertranslation subalgebra $\widehat{\mathfrak{u}}(1)$, yielding the standard $\mathfrak{bms}_3$ algebra.  The hybrid null string arises as a one-parameter deformation of the ILST null string in which the semidirect product action is parameterized by $\lambda$, recovering $\mathfrak{bms}_3$ at $\lambda=1$ and giving a distinct algebra at $\lambda=-1$. These arrows describe relations between constraint algebras and do not imply equivalence of the corresponding physical theories. Omitting a generator changes both the constraint surface and its quotient by the gauge orbits, and is therefore distinct from gauge fixing. In particular, fixing the Carroll--Weyl connection does not eliminate the scaling Gauss-law constraint. The BRST complex of each model must retain the constraints derived from its action before gauge fixing. The last arrow changes the semidirect product action and does not identify the physical state spaces.

We now present the BRST charges, ghost constraint generators, and quantum constraint algebras for each model.

\subsection{The conformal string: \texorpdfstring{$\widehat{\mathfrak{sl}}(2, \mathbb{R}) \rtimes \textit{Vir}$}{sl(2,R)-hat x Virasoro} algebra}
In the Hamiltonian BRST scheme, the classically nilpotent unshifted charge is $Q_{\mathrm c,0}=Q_{\mathrm c,0,\text{matter}}+Q_{\mathrm c,0,\text{ghost}}$:
\begin{align}
Q_{\mathrm c,0} &= \sum_i \sum_n c^i_{-n} M^i_n \nonumber \\
& \quad + \sum_{m,n} \Big[ 2i c^{-1}_{-m} c^1_{-n} b_{0, m+n} - i c^1_{-m} c^0_{-n} b_{1, m+n} + i c^{-1}_{-m} c^0_{-n} b_{-1, m+n} \nonumber \\
& \qquad\qquad - (m+n) c^1_{-m} c^L_{-n} b_{1, m+n} - (m-n) c^{-1}_{-m} c^L_{-n} b_{-1, m+n} \nonumber \\
& \qquad\qquad - m c^0_{-m} c^L_{-n} b_{0, m+n} - \frac{1}{2}(m-n) c^L_{-m} c^L_{-n} b_{L, m+n} \Big],
\end{align}
and quantum charge
\begin{equation}
Q_{\mathrm c}\equiv Q_{\mathrm c,0}-A_0c^0_0-A_Lc^L_0,\qquad A_{-1}=A_1=0.
\end{equation}

The full extended quantum constraints $\tilde{\phi}^i_{\mathrm c, m}$ ($i = -1, 0, 1, L$) are defined through the fundamental BRST anticommutator with the antighosts:
\begin{equation}
\tilde{\phi}^i_{\mathrm c, m} \equiv \{ b_{i, m}, Q_{\mathrm c} \} = M^i_m + G^i_{\mathrm c, m} - A_i \delta_{m,0}.
\end{equation}
Here $A_0$ and $A_L$ are arbitrary normal-ordering zero-mode constants. This uniquely reads off the ghost constraint generators $G^i_{\mathrm c, m} \equiv \{ b_{i, m}, Q_{\mathrm c,0,\text{ghost}} \}$:
\begin{align}
G^{-1}_{\mathrm c, m} &= \sum_{k=-\infty}^{+\infty} \mathopen{:} \left[ 2ic^1_k b_{0, m-k} + ic^0_k b_{-1, m-k} - (m+k)c^L_k b_{-1, m-k} \right] \mathclose{:} \nonumber\\
G^0_{\mathrm c, m} &= \sum_{k=-\infty}^{+\infty} \mathopen{:} \left[ ic^1_k b_{1, m-k} - ic^{-1}_k b_{-1, m-k} - m c^L_k b_{0, m-k} \right] \mathclose{:} \nonumber\\
G^1_{\mathrm c, m} &= - \sum_{k=-\infty}^{+\infty} \mathopen{:} \left[ 2ic^{-1}_k b_{0, m-k} + ic^0_k b_{1, m-k} + (m-k)c^L_k b_{1, m-k} \right] \mathclose{:} \nonumber\\
G^L_{\mathrm c, m} &= - \sum_{k=-\infty}^{+\infty} \mathopen{:} \left[ (k-m)c^1_k b_{1, m-k} + (k+m)c^{-1}_k b_{-1, m-k} + k c^0_k b_{0, m-k} + (k+m)c^L_k b_{L, m-k} \right] \mathclose{:}\nonumber
\end{align}

By substituting the structure constants of the semi-direct product $\widehat{\mathfrak{sl}}(2, \mathbb{R}) \rtimes \textit{Vir}$ algebra, the Jacobi identities consistently restrict the infinite-mode series of the central extensions to the precise polynomial forms expressed in terms of the four anomaly constants:
\begin{align}
d^{1,-1}_m &= 2(i\tilde{d}_4 + i\tilde{d}_3 m), \nonumber\\
d^{L,L}_m &= \tilde{d}_1 m^3 + \tilde{d}_2 m,\nonumber \\
d^{0,L}_m &= \tilde{d}_3 m^2 + \tilde{d}_4 m,\nonumber \\
d^{0,0}_m &= -im\tilde{d}_3.\nonumber
\end{align}
The complete commutation relations of the conformal-string quantum constraint operators $\tilde{\phi}^i_{\mathrm c, m}$ ($i \in \{-1, 0, 1, L\}$) are given by:
\begin{align}
[\tilde{\phi}^1_{\mathrm c, m}, \tilde{\phi}^{-1}_{\mathrm c, n}] &= 2i\tilde{\phi}^0_{\mathrm c, m+n} + 2(i\tilde{d}_4 + i\tilde{d}_3 m)\delta_{m+n,0},\nonumber \\
[\tilde{\phi}^L_{\mathrm c, m}, \tilde{\phi}^L_{\mathrm c, n}] &= (m-n)\tilde{\phi}^L_{\mathrm c, m+n} + (\tilde{d}_1 m^3 + \tilde{d}_2 m)\delta_{m+n,0},\nonumber \\
[\tilde{\phi}^0_{\mathrm c, m}, \tilde{\phi}^L_{\mathrm c, n}] &= m\tilde{\phi}^0_{\mathrm c, m+n} + (\tilde{d}_3 m^2 + \tilde{d}_4 m)\delta_{m+n,0},\nonumber \\
[\tilde{\phi}^0_{\mathrm c, m}, \tilde{\phi}^0_{\mathrm c, n}] &= -im\tilde{d}_3 \delta_{m+n,0},\nonumber \\
[\tilde{\phi}^1_{\mathrm c, m}, \tilde{\phi}^0_{\mathrm c, n}] &= i\tilde{\phi}^1_{\mathrm c, m+n},\nonumber \\
[\tilde{\phi}^{-1}_{\mathrm c, m}, \tilde{\phi}^0_{\mathrm c, n}] &= -i\tilde{\phi}^{-1}_{\mathrm c, m+n},\nonumber \\
[\tilde{\phi}^1_{\mathrm c, m}, \tilde{\phi}^L_{\mathrm c, n}] &= (m+n)\tilde{\phi}^1_{\mathrm c, m+n},\nonumber \\
[\tilde{\phi}^{-1}_{\mathrm c, m}, \tilde{\phi}^L_{\mathrm c, n}] &= (m-n)\tilde{\phi}^{-1}_{\mathrm c, m+n}.\nonumber
\end{align}
The constant coefficients $\tilde{d}_1, \tilde{d}_2, \tilde{d}_3,$ and $\tilde{d}_4$ characterize the four independent central extension branches.  Among these, $\tilde{d}_1$ and $\tilde{d}_3$ are non-trivial cohomology classes (genuine quantum anomalies), while $\tilde{d}_2$ and $\tilde{d}_4$ are trivial coboundaries that can be eliminated by shifting the zero-mode normal-ordering constants $A_L$ and $A_0$, respectively.

\subsection{The ILST null string: \texorpdfstring{$\mathfrak{bms}_3 \cong \widehat{\mathfrak{u}}(1) \rtimes \textit{Vir}$}{bms3 = u(1)-hat x Virasoro} algebra}
To quantize the ILST null string, we construct the classically nilpotent unshifted charge $Q_{\mathrm n,0}=Q_{\mathrm n,0,\text{matter}}+Q_{\mathrm n,0,\text{ghost}}$ over the index sector $i\in\{-1,L\}$:
\begin{equation}
\begin{aligned}
Q_{\mathrm n,0}
&= \sum_n \left( c^L_{-n} M^L_n + c^{-1}_{-n} M^{-1}_n \right) \\
&\quad -\frac12\sum_{m,n}(m-n)c^L_{-m}c^L_{-n}b_{L,m+n} \\
&\quad -\sum_{m,n}(m-n)c^L_{-m}c^{-1}_{-n}b_{-1,m+n},
\end{aligned}
\label{eq:Q_II0_def}
\end{equation}
and quantum charge
\begin{equation}
Q_{\mathrm n}\equiv Q_{\mathrm n,0}-A_Lc^L_0,\qquad A_{-1}=0.
\end{equation}

The full extended quantum constraints $\tilde{\phi}^i_{\mathrm n, m}$ ($i \in \{-1, L\}$) are read off from the anticommutator:
\begin{equation}
\tilde{\phi}^i_{\mathrm n, m} \equiv \{ b_{i, m}, Q_{\mathrm n} \} = M^i_m + G^i_{\mathrm n, m} - A_i \delta_{m,0}
\end{equation}
which defines the BMS ghost constraint generators $G^i_{\mathrm n, m} \equiv \{b_{i, m}, Q_{\mathrm n,0,\text{ghost}}\}$:
\begin{align}
G^{-1}_{\mathrm n, m} &= - \sum_{k=-\infty}^{+\infty} (m+k)
\mathopen{:}c^L_k b_{-1, m-k}\mathclose{:} \nonumber\\
G^L_{\mathrm n, m} &= - \sum_{k=-\infty}^{+\infty} (k+m)
\mathopen{:}c^{-1}_k b_{-1, m-k}\mathclose{:}
- \sum_{k=-\infty}^{+\infty} (k+m)
\mathopen{:}c^L_k b_{L, m-k}\mathclose{:}\nonumber
\end{align}
In fact, the BRST charge and the ghost constraints of the ILST null string can be obtained from those of the conformal string by dropping the $i=0,1$ sectors, retaining only the $i=-1$ and $i=L$ ghost branches.

For the ILST null string, the physical constraint generators are $\tilde{\phi}^{-1}_{\mathrm n, m}$ and $\tilde{\phi}^L_{\mathrm n, m}$. The quantum constraint algebra corresponds to the $\mathfrak{bms}_3$ algebra with potential central extensions. The Jacobi identities of this two-generator algebra restrict the central extensions to:
\begin{align}
d^{L,L}_m &= \tilde{d}_1 m^3 + \tilde{d}_2 m, \nonumber\\
d^{L,-1}_m &= \tilde{d}_5 m^3 + \tilde{d}_6 m,\nonumber \\
d^{-1,-1}_m &= 0.
\end{align}
The complete quantum commutation relations of the constraints $\tilde{\phi}^i_{\mathrm n, m}$ ($i \in \{-1, L\}$) are given by:
\begin{align}
[\tilde{\phi}^L_{\mathrm n, m}, \tilde{\phi}^L_{\mathrm n, n}] &= (m-n)\tilde{\phi}^L_{\mathrm n, m+n} + (\tilde{d}_1 m^3 + \tilde{d}_2 m)\delta_{m+n,0},\nonumber \\
[\tilde{\phi}^L_{\mathrm n, m}, \tilde{\phi}^{-1}_{\mathrm n, n}] &= (m-n)\tilde{\phi}^{-1}_{\mathrm n, m+n} + (\tilde{d}_5 m^3 + \tilde{d}_6 m)\delta_{m+n,0},\nonumber \\
[\tilde{\phi}^{-1}_{\mathrm n, m}, \tilde{\phi}^{-1}_{\mathrm n, n}] &= 0,\nonumber
\end{align}
where the coefficients $\tilde{d}_1, \tilde{d}_2, \tilde{d}_5,$ and $\tilde{d}_6$ are the central extension parameters of the $\mathfrak{bms}_3$ algebra.  Among these, $\tilde{d}_1$ and $\tilde{d}_5$ are non-trivial cohomology classes (genuine quantum anomalies), while $\tilde{d}_2$ and $\tilde{d}_6$ are trivial coboundaries that can be eliminated by shifting the zero-mode normal-ordering constants $A_L$ and $A_{-1}$, respectively.
This statement classifies the unrestricted algebra: in the representative \(A_{-1}=0\) adopted immediately after Eq.~\eqref{eq:Q_II0_def}, direct vacuum evaluation gives \(\tilde d_6=0\).

\subsection{The Carroll--Weyl gauged string: \texorpdfstring{Carroll--Weyl $\cong \widehat{\mathfrak{aff}}(1) \rtimes \textit{Vir}$}{Carroll-Weyl = aff(1)-hat x Virasoro} algebra}
For the Carroll--Weyl gauged string, the classically nilpotent unshifted charge $Q_{\mathrm g,0}=Q_{\mathrm g,0,\text{matter}}+Q_{\mathrm g,0,\text{ghost}}$ is constructed over $i\in\{-1,0,L\}$:
\begin{align}
Q_{\mathrm g,0} &= \sum_n \left( c^{-1}_{-n} M^{-1}_n + c^L_{-n} M^L_n + c^0_{-n} M^0_n \right) \nonumber \\
& \quad + \sum_{m,n} \Big[ i c^{-1}_{-m} c^0_{-n} b_{-1, m+n} - (m-n) c^{-1}_{-m} c^L_{-n} b_{-1, m+n} \nonumber \\
& \qquad\qquad\quad - m c^0_{-m} c^L_{-n} b_{0, m+n} - \frac{1}{2}(m-n) c^L_{-m} c^L_{-n} b_{L, m+n} \Big],
\end{align}
and quantum charge is 
\begin{equation}
Q_{\mathrm g}\equiv Q_{\mathrm g,0}-A_0c^0_0-A_Lc^L_0,\qquad A_{-1}=0.
\end{equation}

The full extended constraints are defined via the BRST anticommutators:
\begin{equation}
\tilde{\phi}^i_{\mathrm g, m} \equiv \{ b_{i, m}, Q_{\mathrm g} \} = M^i_m + G^i_{\mathrm g, m} - A_i \delta_{m,0}
\end{equation}
yielding the ghost constraint generators $G^i_{\mathrm g, m} \equiv \{b_{i, m}, Q_{\mathrm g,0,\text{ghost}}\}$ in the unified index conventions:
\begin{align}
G^{-1}_{\mathrm g, m} &= \sum_{k=-\infty}^{+\infty} \mathopen{:} \left[ ic^0_k b_{-1, m-k} - (m+k)c^L_k b_{-1, m-k} \right] \mathclose{:} \nonumber\\
G^0_{\mathrm g, m} &= \sum_{k=-\infty}^{+\infty} \mathopen{:} \left[ -ic^{-1}_k b_{-1, m-k} - m c^L_k b_{0, m-k} \right] \mathclose{:} \nonumber\\
G^L_{\mathrm g, m} &= - \sum_{k=-\infty}^{+\infty} \mathopen{:} \left[ (k+m)c^{-1}_k b_{-1, m-k} + k c^0_k b_{0, m-k} + (k+m)c^L_k b_{L, m-k} \right] \mathclose{:}\nonumber
\end{align}
In fact, the BRST charge and the ghost constraints of the Carroll--Weyl gauged string can be obtained from those of the conformal string by dropping the $i=1$ sector, retaining the $i=-1$, $i=0$, and $i=L$ ghost branches.

For the Carroll--Weyl gauged string, the constraint generators are $\tilde{\phi}^{-1}_{\mathrm g, m}$, $\tilde{\phi}^0_{\mathrm g, m}$, and $\tilde{\phi}^L_{\mathrm g, m}$, which form a subalgebra of the conformal string constraints. Compared to the conformal string, because the special conformal constraint generator $\tilde{\phi}^1$ is excluded, the Jacobi identity does not require the central parameter of $[\tilde{\phi}^0, \tilde{\phi}^0]$ to equal that of $[\tilde{\phi}^0, \tilde{\phi}^L]$. Consequently, these two coefficients decouple, denoted by $\tilde{d}_3$ and $\tilde{d}'_3$, respectively. The Jacobi identities of this subalgebra restrict the central extensions to:
\begin{align}
d^{L,L}_m &= \tilde{d}_1 m^3 + \tilde{d}_2 m,\nonumber \\
d^{0,L}_m &= \tilde{d}'_3 m^2 + \tilde{d}_4 m,\nonumber \\
d^{0,0}_m &= -im\tilde{d}_3.\nonumber
\end{align}
The complete quantum commutation relations for the Carroll--Weyl gauged string constraints are then:
\begin{align}
[\tilde{\phi}^L_{\mathrm g, m}, \tilde{\phi}^L_{\mathrm g, n}] &= (m-n)\tilde{\phi}^L_{\mathrm g, m+n} + (\tilde{d}_1 m^3 + \tilde{d}_2 m)\delta_{m+n,0},\nonumber \\
[\tilde{\phi}^0_{\mathrm g, m}, \tilde{\phi}^L_{\mathrm g, n}] &= m\tilde{\phi}^0_{\mathrm g, m+n} + (\tilde{d}'_3 m^2 + \tilde{d}_4 m)\delta_{m+n,0},\nonumber \\
[\tilde{\phi}^0_{\mathrm g, m}, \tilde{\phi}^0_{\mathrm g, n}] &= -im\tilde{d}_3 \delta_{m+n,0},\nonumber \\
[\tilde{\phi}^{-1}_{\mathrm g, m}, \tilde{\phi}^0_{\mathrm g, n}] &= -i\tilde{\phi}^{-1}_{\mathrm g, m+n},\nonumber \\
[\tilde{\phi}^{-1}_{\mathrm g, m}, \tilde{\phi}^L_{\mathrm g, n}] &= (m-n)\tilde{\phi}^{-1}_{\mathrm g, m+n},\nonumber \\
[\tilde{\phi}^{-1}_{\mathrm g, m}, \tilde{\phi}^{-1}_{\mathrm g, n}] &= 0.\nonumber
\end{align}
The five independent coefficients $\tilde{d}_1, \tilde{d}_2, \tilde{d}_3, \tilde{d}'_3,$ and $\tilde{d}_4$ characterize the central extensions of the Carroll-Weyl gauged string.  Among these, $\tilde{d}_1$, $\tilde{d}_3$, and $\tilde{d}'_3$ are non-trivial cohomology classes (genuine quantum anomalies), while $\tilde{d}_2$ and $\tilde{d}_4$ are trivial coboundaries that can be eliminated by shifting the zero-mode normal-ordering constants $A_L$ and $A_0$, respectively.

\subsection{The hybrid null string: \texorpdfstring{$\mathfrak{g}^{(1)}_\lambda \cong \widehat{\mathfrak{u}}(1) \rtimes_\lambda \textit{Vir}$}{g(1)-lambda = u(1)-hat x Virasoro} algebra}
In this model, the classically nilpotent unshifted charge $Q_{\mathrm h,0}=Q_{\mathrm h,0,\text{matter}}+Q_{\mathrm h,0,\text{ghost}}$ is defined over $i\in\{-1,L_\lambda\}$ as:
\begin{align}
Q_{\mathrm h,0} &= \sum_n \left( c^{L_\lambda}_{-n} M^{L_\lambda}_n + c^{-1}_{-n} M^{-1}_n \right) \nonumber \\
& \quad - \sum_{m,n} \left[ \frac{1}{2}(m-n) c^{L_\lambda}_{-m} c^{L_\lambda}_{-n} b_{L_\lambda, m+n} + (\lambda m - n) c^{L_\lambda}_{-m} c^{-1}_{-n} b_{-1, m+n} \right],
\end{align}
and quantum charge is
\begin{equation}
Q_{\mathrm h}\equiv Q_{\mathrm h,0}-A_{L_\lambda}c^{L_\lambda}_0,\qquad A_{-1}=0\,.
\end{equation}

The full hybrid extended constraints are read off via the BRST anticommutators:
\begin{equation}
\tilde{\phi}^i_{\mathrm h, m} \equiv \{ b_{i, m}, Q_{\mathrm h} \} = M^i_m + G^i_{\mathrm h, m} - A_i \delta_{m,0}
\end{equation}
yielding the hybrid ghost generators $G^i_{\mathrm h, m} \equiv \{b_{i, m}, Q_{\mathrm h,0,\text{ghost}}\}$:
\begin{align}
G^{-1}_{\mathrm h, m} &= - \sum_{k=-\infty}^{+\infty} [ \lambda k + m ] \mathopen{:} c^{L_\lambda}_k b_{-1, m-k} \mathclose{:} \nonumber\\
G^{L_\lambda}_{\mathrm h, m} &= - \sum_{k=-\infty}^{+\infty} [ \lambda m + k ] \mathopen{:} c^{-1}_k b_{-1, m-k} \mathclose{:} - \sum_{k=-\infty}^{+\infty} (k+m) \mathopen{:} c^{L_\lambda}_k b_{L_\lambda, m-k} \mathclose{:}\nonumber
\end{align}

For the hybrid null string, the complete centrally extended commutators can be written uniformly as
\begin{align}
[\tilde\phi^{L_\lambda}_{\mathrm h,m},\tilde\phi^{L_\lambda}_{\mathrm h,n}] &=(m-n)\tilde\phi^{L_\lambda}_{\mathrm h,m+n}+(\tilde d_1m^3+\tilde d_2m)\delta_{m+n,0}, \nonumber\\
[\tilde\phi^{L_\lambda}_{\mathrm h,m},\tilde\phi^{-1}_{\mathrm h,n}] &=(\lambda m-n)\tilde\phi^{-1}_{\mathrm h,m+n}+g_\lambda(m)\delta_{m+n,0}, \nonumber\\
[\tilde\phi^{-1}_{\mathrm h,m},\tilde\phi^{-1}_{\mathrm h,n}] &=h_\lambda(m)\delta_{m+n,0}.
\label{eq:quantum_algebra_IV}
\end{align}
Here we write the possible central extensions involving $\tilde{\phi}^{-1}_m$ as $g_\lambda(m)\delta_{m+n,0}$ and $h_\lambda(m)\delta_{m+n,0}$.  The \((L_\lambda,L_\lambda,-1)\) and \((L_\lambda,-1,-1)\) Jacobi identities give the independent recurrences
\begin{align}
\bigl(m+(\lambda+1)n\bigr)g_\lambda(m)-\bigl((\lambda+1)m+n\bigr)g_\lambda(n)-(m-n)g_\lambda(m+n) &=0, \nonumber\\
\bigl(\lambda m+(\lambda+1)n\bigr)h_\lambda(m)-\bigl((\lambda+1)m+\lambda n\bigr)h_\lambda(n) &=0.\nonumber
\end{align}
Within polynomials of degree at most three, the complete solutions are: 
\begin{itemize}
    \item For generic \(\lambda\notin\{-1,0,1\}\), \(g_\lambda(m)=\gamma_1m\) and \(h_\lambda(m)=0\);
    \item At \(\lambda=-1\), \(g_{-1}(m)=\gamma_0+\gamma_1m\) and \(h_{-1}(m)=0\);
    \item At \(\lambda=0\), \(g_0(m)=\gamma_1m+\gamma_2m^2\) and \(h_0(m)=\kappa_1m\);
    \item At \(\lambda=1\), \(g_1(m)=\gamma_1m+\gamma_3m^3\) and \(h_1(m)=0\).
\end{itemize}
The Virasoro-sector cocycle remains $d^{L_\lambda,L_\lambda}_m=\tilde d_1m^3+\tilde d_2m$. Here \(\gamma_r,\kappa_1\in\mathbb C\) are arbitrary constants. These cocycle classes agree with the known central extensions of the \(W(a,b)\) family~\cite{Gao:2011Wab,Figueroa-OFarrill:2024wgs}, upon identifying \(a=0\) and \(b=-\lambda\) (equivalently, \(\lambda_{\rm there}=-\lambda\) in Ref.~\cite{Figueroa-OFarrill:2024wgs}).

A zero-mode shift \(\tilde\phi^{-1}_0\mapsto\tilde\phi^{-1}_0+s\) changes the cocycle by \(g_\lambda(m)\mapsto g_\lambda(m)-(\lambda+1)sm\).  Hence the linear terms $\gamma_1 m$ at generic $\lambda\neq -1$ are coboundaries; at $\lambda=-1$ the shift vanishes, so both the constant $\gamma_0$ and linear term $\gamma_1 m$ are non-trivial cocycles.  The additional non-trivial cocycles are $\gamma_2 m^2$ and $\kappa_1 m$ at $\lambda=0$, and the familiar $\gamma_3 m^3$ at $\lambda=1$.  As before, $\tilde d_1$ is non-trivial while $\tilde d_2$ is removed by the \(L_{\lambda,0}\) normal-ordering zero-mode constant.  The subsequent vacuum calculations determine which of these allowed classes are actually realized; in the representation studied in this paper, all central extensions involving $\tilde{\phi}^{-1}_m$ evaluate to zero.

\section{Vacuum structures and quantum anomalies}
\label{sec:anomalies}

In this section, we analyze the quantum anomalies $d^{ij}_m$ for two choices of vacuum, induced and flipped, with the constraints and ghost system specified separately for each model. We do not assume that the ILST vacuum classification exhausts the possible vacua of the other theories. The vacuum conditions determine the normal-ordering prescription, while physical states must separately satisfy the BRST conditions. For the flipped vacuum, only finitely many terms in the normal-ordered BRST charge contribute on states of finite oscillator level with polynomial dependence on the zero modes. All central terms, including zero-mode coboundaries, must cancel. A failure of BRST nilpotency therefore excludes the chosen representation, without ruling out other quantizations of the theory. In particular, the light-cone quantization of Ref.~\cite{Rasulian:2026lc} uses a constant vacuum wavefunctional in the Schr\"odinger representation; see Eqs.~(3.10)--(3.12) there. It is annihilated by the independent transverse momenta but not by the coordinate operators. For the nonzero transverse modes, these conditions agree with those of the induced vacuum. The absence of a critical dimension in that quantization is consistent with our formal induced-vacuum result. Since that construction fixes $\bar p^+\ne0$ and solves the local constraints before quantization, equivalence with the covariant induced prescription requires a separate comparison of the physical state spaces.

As reviewed in Appendix~\ref{app:wick}, by using Wick's theorem  the commutator of two normal-ordered constraint generators can be expressed as a sum over contractions.  For the bilinear generators used here, the single-contraction terms reproduce the classical constraint algebra, while the fully contracted term contains double contractions of two pairings, which gives rise to a pure number. Consequently, such double contractions leads to a central extension of the quantum constraint algebra and may obstruct BRST nilpotency.  The two vacuum choices lead to different contractions and hence to different quantum anomalies.

\subsection{The induced vacuum}
The universal part of the \textbf{induced vacuum} is the momentum condition
\begin{equation}
p^M_n |0\rangle_{\text{Ind}} = 0 \qquad (\forall n \in \mathbb{Z}).
\end{equation}
Here $M$ runs over $D+2$ target coordinates for the conformal string and $D$ for the other models. We use $x_n^\dagger=x_{-n}$, $p_n^\dagger=p_{-n}$, $(c_n^i)^\dagger=c_{-n}^i$, and $b_{i,n}^\dagger=b_{i,-n}$, and take the Hermitian Weyl-ordered charge as the reference for operator ordering. In the induced vacuum of the conformal string, the ghost variables satisfy $b_{-1,n}|0\rangle=c_n^1|0\rangle=0$; the four choices of ghost fixing are listed in Appendix~\ref{app:induced}, while the other models impose antighost annihilation for every ghost pair. The notation $\langle A\rangle_{\rm Ind}$, including the Wick notation below, denotes the matrix element $\langle0^\vee|A|0\rangle$ between the vacuum and its dual vacuum state, with $\langle0^\vee|0\rangle=1$. The dual satisfies $\langle0^\vee|x_n=0$, $\langle0^\vee|c_n^i=0$ for a $B_i$ choice, and $\langle0^\vee|b_{i,n}=0$ for a $C_i$ choice. The canonical (anti)commutation relations define a nondegenerate pairing on finite polynomials; $\langle0^\vee|$ is distinct from the Hermitian conjugate of $|0\rangle$. Following Ref.~\cite{Hwang:1998brst}, the $\alpha=0$ prescription uses the same even regulator $f_s(0)=1$, $f_s(n)=|n|^{-s}$ for $n\ne0$ in the matter and ghost basic (anti)commutators, retaining the ordering terms before analytic continuation. The single and shifted double sums have zero finite part. For each ghost choice of the conformal string, the calculated terms in the BRST consistency conditions for the vacuum and its dual are proportional to $(D-2)S_\alpha$ and vanish at $\alpha=0$. The nonzero-$\alpha$ anomalies and Jacobi-identity violations displayed in Appendix~\ref{app:induced} give individual contributions from contractions and operator ordering; they do not determine the complete Hermitian regulated operator algebra. The resulting operator equations are formal and their physical interpretation remains an open questions.\subsection{The flipped vacuum}
The \textbf{flipped vacuum} $|0\rangle_{\text{Flip}}$ is a highest-weight representation vacuum where positive frequency modes of coordinates, momenta, and ghosts act as annihilation operators:
\begin{equation}
x^M_n |0\rangle_{\text{Flip}} = p^M_n |0\rangle_{\text{Flip}} = b_{i, n} |0\rangle_{\text{Flip}} = c^i_n |0\rangle_{\text{Flip}} = 0 \quad (\forall n > 0)
\end{equation}
with the zero modes of the momentum satisfying $p^M_0 |0\rangle_{\text{Flip}} = 0$. Additionally, to resolve the degeneracy of the ghost zero modes, we manually prescribe that the antighost zero modes annihilate the vacuum:
\begin{equation}
b_{i,0} |0\rangle_{\text{Flip}} = 0
\end{equation}
which consequently implies $c^i_0 |0\rangle_{\text{Flip}} \neq 0$ due to the anticommutation relation $\{b_{i,0}, c^j_0\} = \delta_i^j$. Normal-ordered products of zero-mode ghost operators are then defined with $b_{i,0}$ placed to the right of $c^j_0$.

The finite double contractions are evaluated in Appendix~\ref{app:flipped_derivations}.  Table~\ref{tab:critical_dims} collects the non-trivial quantum anomalies; none of the model zero-mode coboundaries---including \(\tilde d_2,\tilde d_4,\tilde d_6\) and the removable linear part of \(g_\lambda(m)\)---contributes to the consistency conditions on the critical dimension. Thus we do not show them in Table~\ref{tab:critical_dims}.
\begin{table}[t!]
\centering\footnotesize
\caption{Non-trivial flipped-vacuum anomalies and their consistency conditions.  All displayed classes must vanish simultaneously.}
\label{tab:critical_dims}
\setlength{\tabcolsep}{4pt}
\resizebox{\textwidth}{!}{%
\begin{tabular}{lccc}
\toprule
String & \(\tilde d_1\) (Virasoro) & Scaling anomalies & Consistent \(D\)\\
\midrule
Conformal string
& \(\dfrac{D-26}{6}\)
& \(\tilde d_3=i\bigl(\frac{D+2}{4}-2\bigr)\)
& none (\(26\) vs.\ \(6\))\\[2pt]
ILST null string
& \(\dfrac{D-26}{6}\)
& ---
& \(D=26\)\\[2pt]
Carroll--Weyl gauged string
& \(\dfrac{D-27}{6}\)
& \(\tilde d_3=i\bigl(\frac D4-1\bigr),\quad
   \tilde d'_3=i\bigl(\frac D4-\frac32\bigr)\)
& none (\(27\) vs.\ \(4\) vs.\ \(6\))\\[2pt]
Hybrid null string
& \(\dfrac{D(3\lambda^2-1)-2(6\lambda^2+6\lambda+14)}{12}\)
& ---
& \(D=D(\lambda)\)\\
\bottomrule
\end{tabular}
}
\end{table}

The conformal and Carroll--Weyl gauged strings are therefore structurally anomalous: their Virasoro and scaling classes impose incompatible conditions on dimensions.  The ILST null string has no additional non-zero quantum anomaly and reproduces \(D=26\).  In the hybrid null string,
\begin{equation}
D(\lambda)
=\frac{2(6\lambda^2+6\lambda+14)}{3\lambda^2-1},
\label{eq:hybrid_critical_dimension}
\end{equation}
with \(D(1)=26\).  Equivalently, a prescribed dimension obeys
\begin{equation}
(3D-12)\lambda^2-12\lambda-(D+28)=0.
\end{equation}
For this algebraic range analysis \(\lambda\) is first treated as a real parameter; physical target dimensions are subsequently restricted to positive integers.  Its discriminant is \(12(D^2+24D-100)\).  Thus a real \(\lambda\) exists precisely when
\(D\leq-12-2\sqrt{61}\) or \(D\geq-12+2\sqrt{61}\), with the poles \(3\lambda^2=1\) excluded from Eq.~\eqref{eq:hybrid_critical_dimension}.  In particular, every integer \(D\geq4\) is realized; the smallest such dimension is \(D=4\), for which \(\lambda=-8/3\) and Eq.~\eqref{eq:lambda_Delta} gives \(\Delta=-11/5\).  If both \(\lambda\) and \(D\geq4\) are required to be integers, the only solutions are \((\lambda,D)=(1,26)\) and \((-1,14)\).  The latter is a point not covered by finite \(\Delta\) in Eq.~\eqref{eq:lambda_Delta}.  Direct evaluation also sets every exceptional quantum anomaly $g_{\lambda}(m)$ and $h_{\lambda}(m)$ at \(\lambda=-1,0,1\) to zero, so these extra Jacobi-compatible classes do not modify Eq.~\eqref{eq:hybrid_critical_dimension}.

\section{Conclusion}
\label{sec:conclusion}

We have formulated and quantized four tensionless bosonic string theories in a Hamiltonian BRST framework. Expressing the fields, constraints, and ghosts in terms of modes makes it possible to separate the universal matter dynamics from the model-dependent residual gauge structure and to compare their quantum anomalies on the same footing.

\medskip
\noindent\textbf{Summary of the four strings.}
The four formulations of null bosonic strings and their residual constraint algebras are summarized as follows:
\begin{center}
\small
\setlength{\tabcolsep}{5pt}
\begin{tabular}{llc}
\toprule
String & Constraint algebra & Ghost branches \\
\midrule
Conformal string
& \(\widehat{\mathfrak{sl}}(2,\mathbb R)\rtimes\textit{Vir}\)
& \(4\) \\
ILST null string
& \(\mathfrak{bms}_3\cong\widehat{\mathfrak u}(1)\rtimes\textit{Vir}\)
& \(2\) \\
Carroll--Weyl gauged string
& Carroll--Weyl \(\cong\widehat{\mathfrak{aff}}(1)\rtimes\textit{Vir}\)
& \(3\) \\
Hybrid null string
& \(\mathfrak g^{(1)}_\lambda\cong\widehat{\mathfrak u}(1)\rtimes_\lambda\textit{Vir}\)
& \(2\) \\
\bottomrule
\end{tabular}
\end{center}
The ILST null, Carroll--Weyl gauged, and hybrid null strings have a \(D\)-dimensional target space, whereas the conformal string uses a \((D+2)\)-dimensional embedding that makes target-space conformal symmetry manifest.  All four theories reduce after gauge fixing to the same free matter action,
\(
S=(2\pi)^{-1}\int d^2\sigma\,(\partial_\tau X)^2/2,
\).
However, they differ in their residual constraint sets and hence in the number and couplings of their ghost branches.  The conformal string contains the full \(\widehat{\mathfrak{sl}}(2,\mathbb R)\rtimes\textit{Vir}\) algebra; removing the special-conformal and then the scaling branch gives the Carroll--Weyl gauged and ILST null strings, while the hybrid null string deforms the Virasoro action on the worldsheet-supertranslation sector of the ILST null string. 

Our BRST analysis makes this distinction precise. For each string , the unshifted charge \(Q_{s,0}\) (\(s\in\{\mathrm c,\mathrm n,\mathrm g,\mathrm h\}\)) and the constraints \(H^i_m\) were kept separate from the zero-mode-shifted charge \(Q_s\) and constraints \(\tilde\phi^i_m\).  The Jacobi identities determine the allowed two-cocycles; zero-mode redefinitions remove the coboundaries, while every remaining cohomology class must vanish for \(Q_s^2=0\). In particular, the hybrid null string admits exceptional Jacobi-compatible cocycles at \(\lambda=-1,0,1\), although they vanish in the vacuum representations studied here.

\medskip
\noindent\textbf{Vacuum dependence of the quantum anomalies.}
The consistency conditions depend decisively on the vacuum. In the induced vacuum, BRST compatibility fixes a non-standard ghost field annihilation condition for the conformal string. The quantum anomaly fails the cocycle condition and is therefore neither a physical central extension nor a removable coboundary.  Within the generalized-zeta family used here, the symmetric \(\alpha=0\) prescription removes this obstruction.  All non-trivial induced-vacuum coefficients then vanish in the four strings, while the remaining zero-mode terms are removable by normal-ordering constants.  Consequently this regulated induced scheme selects no critical dimension.

The flipped highest-weight vacuum gives  qualitatively different results.  Its quantum anomalies are given in Table~\ref{tab:critical_dims}.  The conformal string is structurally anomalous because its Virasoro and scaling anomalies vanish at the incompatible dimensions \(D=26\) and \(D=6\).  The Carroll--Weyl gauged string has three incompatible conditions, \(D=27\), \(D=4\), and \(D=6\), from its Virasoro and two scaling anomalies.  Since these are independent cohomology classes, zero-mode shifts cannot reconcile them.  The ILST null string has no additional non-zero central extension and is consistent at \(D=26\), reproducing the familiar bosonic-string critical dimension within the tensionless BRST algebra.

For the hybrid null string, the flipped-vacuum condition is the continuous family
\[
D(\lambda)=\frac{2(6\lambda^2+6\lambda+14)}{3\lambda^2-1}.
\]
It contains \(D(1)=26\), where the algebra reduces to \(\mathfrak{bms}_3\). If $\lambda$ is real,  the critical dimension $D$ can take any integer value $D\geq 4$ from the above relation. The smallest case is \(D=4\) at \(\lambda=-8/3\), corresponding to \(\Delta=-11/5\).  If both \(D\geq4\) and \(\lambda\) are restricted to integers, only \((\lambda,D)=(1,26)\) and \((-1,14)\) remain; the latter is a point outside the finite-\(\Delta\) Stueckelberg parametrization.  Direct evaluation sets all exceptional quantum anomalies at \(\lambda=-1,0,1\) to zero, so the extra allowed cocycles do not modify this critical-dimension curve.

\medskip
\noindent\textbf{Spacetime conformal symmetry in the ILST null string.}
The ILST null string also exhibits target-space conformal symmetry in addition to its worldsheet symmetries. In the induced vacuum and the \(\alpha=0\) prescription, the ghost-completed generators \(\mathsf P^\mu,\mathsf M^{\mu\nu},\mathcal D,K^\mu\) are \(Q_{\mathrm n}\)-closed and obey the \(SO(D,2)\) algebra. In contrast, in the flipped vacuum, Eq.~\eqref{eq:A_mu_explicit} contains two independent operator structures, \(c^L_{-n}x^\mu_n\) and \(c^{-1}_{-n}p^\mu_n\), whose matter and ghost coefficients have incompatible mode dependence, and no value of \(D\) makes both structures vanish simultaneously.  Hence the special-conformal charge is not \(Q_{\mathrm n}\)-closed even at \(D=26\), where the worldsheet BRST charge itself can be nilpotent.  The cancellation of the worldsheet gauge anomaly is therefore necessary but not sufficient to protect the global target-space symmetry that replaces Weyl invariance.

This contrast reflects the different status of target-space conformal symmetry in the conformal string and the ILST null string. In the conformal string, $SO(D,2)$ is manifest as the Lorentz symmetry of the $(D+2)$-dimensional embedding, and the additional $X^2$ and $X\cdot P$ constraints enlarge the worldsheet BRST algebra. In the flipped vacuum, independent central extensions of this enlarged algebra require $D=26$ and $D=6$, respectively, so no dimension yields a nilpotent BRST charge. In the ILST null string, by contrast, $SO(D,2)$ is realized by composite global charges and is not part of the $\mathfrak{bms}_3$ gauge-constraint algebra. The $\mathfrak{bms}_3$ symmetry is anomaly free at $D=26$, whereas the fixed normal-ordered representative of $K^\mu$ is not $Q_{\mathrm n}$-closed for any $D$. Thus the worldsheet BRST consistency and target-space conformal invariance are distinct quantum conditions in the ILST null string: even though at $D=26$ the former is satisfied,  the latter fails for the representative studied here.

\medskip
\noindent\textbf{Outlook.}
There are several directions worthy of further investigation.   First, supersymmetric versions of the four strings could be analyzed within the same Hamiltonian BRST framework.  The additional matter fermions and superghosts would modify the quantum anomalies, so it is important to determine whether the incompatible anomaly conditions of the conformal and Carroll--Weyl gauged strings persist and how the hybrid null string curve \(D(\lambda)\) is altered.

Second, if we want to understand the null string better, the flipped-vacuum analysis should be extended beyond the  BRST nilpotency. For the anomaly-free candidates—the ILST null string at $D=26$ and the candidate points on the hybrid null string curve $D(\lambda)$—one should compute the BRST cohomology, extract the physical-state spectrum, and construct BRST-invariant vertex operators before assessing whether they define non-trivial, consistent string theories.  These ingredients would also provide the starting point for computing scattering amplitudes and for testing how the flipped-vacuum breaking of the ILST null string \(SO(D,2)\) charge algebra is reflected in physical observables.

Finally, the hybrid null string raises questions that do not occur in the other three theories.  The continuous family \(D(\lambda)\), the exceptional cocycles at \(\lambda=-1,0,1\), and the fact that \(\lambda=-1\) lies outside the finite-\(\Delta\) Stueckelberg parametrization deserve a more direct physical interpretation.  In particular, it remains to determine which points on the critical dimension curve \eqref{eq:hybrid_critical_dimension} admit a non-trivial BRST spectrum and whether different values of \(\lambda\) define inequivalent quantum theories.

\acknowledgments

The authors thank Sarthak Duary and Sourav Maji for valuable discussions.  This research was supported in part by the National Natural Science Foundation of China (NSFC) under Grants  No.~12275004, No.~12588101.

\appendix
\section{Wick contractions and normal-ordered commutators}
\label{app:wick}
\noindent\textbf{Normal ordering.}
Consider the free Fourier modes \(x^M_m,p^M_n\) and the ghost modes \(c^i_m,b_{i,n}\). Colons \(\mathopen{:}\cdots\mathclose{:}\) denote normal ordering with respect to the vacuum of the relevant subsection, which moves annihilation operators to the right of creation operators, with one minus sign for each interchange of two fermionic fields. Which modes are annihilators is fixed by the choice of vacuum. In particular,
\begin{equation}
x^M_{m>0}|0\rangle_{\rm Flip}
=p^M_{m>0}|0\rangle_{\rm Flip}=0,
\qquad
p^M_n|0\rangle_{\mathrm{Ind}}=0\quad(n\in\mathbb Z).
\label{eq:appA_vacuum_examples}
\end{equation}
The ghost sector follows the branch-dependent prescriptions described in Appendices~\ref{app:induced} and~\ref{app:flipped_derivations}. The vacuum expectation value of a non-empty normal-ordered product vanishes.

\medskip
\noindent\textbf{Wick's theorem and the commutator.}
Let \(X=\mathopen{:}A_1\cdots A_r\mathclose{:}\) and \(Y=\mathopen{:}B_1\cdots B_s\mathclose{:}\) be non-empty products of free fields, each already being normal ordered.  Assign every elementary field a Grassmann parity \(|A|\in\{0,1\}\) and write the graded commutator as
\begin{equation}
[A,B\}\equiv AB-(-1)^{|A||B|}BA,
\label{eq:graded_commutator_appA}
\end{equation}
which reduces to the ordinary commutator whenever at least one operator is Grassmann even and to the anticommutator when both are odd.  Wick's theorem gives
\begin{equation}
\begin{aligned}
\mathopen{:}X\mathclose{:}\,\mathopen{:}Y\mathclose{:}
={}&\mathopen{:}XY\mathclose{:}
+\sum_{\text{one contraction}}(\pm)\,
\contraction{}{A}{}{B\quad}A_{i_1}B_{j_1}\,\mathopen{:}\text{rest}\mathclose{:}\\
&+\sum_{\text{two contractions}}(\pm)\,
\contraction{}{A}{}{B\quad}A_{i_1}B_{j_1}\,
\contraction{}{A}{}{B\quad}A_{i_2}B_{j_2}\,\mathopen{:}\text{rest}\mathclose{:}
+\cdots .
\end{aligned}
\label{eq:wick_product}
\end{equation}
Here
\begin{equation}
\contraction{}{A}{}{B}AB
\equiv AB-\mathopen{:}AB\mathclose{:}
=\langle0|AB|0\rangle
\label{eq:contraction_def}
\end{equation}
is the contraction of two elementary free fields.  The sign in each term of Eq.~\eqref{eq:wick_product} is \((-1)^P\), where \(P\) counts the fermionic transpositions required to bring the contracted fields next to one another.  There are no contractions internal to \(X\) or \(Y\), because each factor was normal ordered before the product was formed, and the expansion contains at most \(\min(r,s)\) contractions.

For simplicity, the ordinary-commutator example below is restricted to the case in which at least one of \(X\) and \(Y\) is Grassmann even.  In this case the graded bracket reduces to the ordinary commutator.  Applying Eq.~\eqref{eq:wick_product} to both orderings gives
\begin{equation}
\begin{aligned}
\relax[\mathopen{:}X\mathclose{:},\mathopen{:}Y\mathclose{:}]
={}&\bigl(\mathopen{:}XY\mathclose{:}-\mathopen{:}YX\mathclose{:}\bigr)\\
&+\bigl(\text{contraction terms from }\mathopen{:}X\mathclose{:}\mathopen{:}Y\mathclose{:}
-\text{contraction terms from }\mathopen{:}Y\mathclose{:}\mathopen{:}X\mathclose{:}\bigr).
\end{aligned}
\label{eq:wick_commutator}
\end{equation}
Under this parity restriction, the fully normal-ordered terms cancel.  The commutator is therefore organized entirely by the number of contractions:
\begin{itemize}
\item \textbf{Single contraction.}  One field from \(X\) and one from \(Y\) are paired, leaving \(r+s-2\) fields inside the normal-ordering symbol.  The difference between the two operator orderings yields the ordinary classical commutators.
\item \textbf{Multiple contractions.}  Contracting \(n\) pairs leaves \(r+s-2n\) fields.  If fields remain, the contribution is still operator valued, which happens in Appendix~\ref{app:SO(D,2)}.  If all fields are paired, the result is a pure c-number.  Since every constraint generator in this paper is a sum of bilinears, its fully contracted commutator contains exactly two contractions.  These double contractions give the central extensions computed in Appendices~\ref{app:induced} and~\ref{app:flipped_derivations}.
\end{itemize}
This is the same mechanism by which double contractions of Virasoro generators produce the central charge in the tensile bosonic string.  Here the ghost bilinears add fermionic signs, and the resulting cancellation or mismatch between matter and ghost contributions determines the anomaly conditions.

\smallskip
\noindent\textbf{Double-contraction notation.}
\begin{itemize}
\item \(\langle\cdots\rangle\) denotes a vacuum expectation value, including the two-point function of elementary fields; it is a c-number.
\item \(\langle\!\langle\cdots\rangle\!\rangle\) denotes extraction of the double-contraction contribution from a product or commutator of composite operators.  It is a pure c-number when all fields are paired, but it is an operator---a c-number coefficient multiplying the uncontracted fields---when fields remain.
\end{itemize}
The second notation is needed in the following appendices. In Appendix~\ref{app:SO(D,2)}, \(K^\mu\) is cubic in matter modes, so its double contraction with a BRST charge leaves one matter mode and one ghost mode uncontracted.  Appendices~\ref{app:SO(D,2)}, \ref{app:induced}, and~\ref{app:flipped_derivations} apply these rules to the two vacuum prescriptions without introducing any additional contraction convention.
\section{Derivation of world-sheet quantum anomaly in the induced vacuum}
\label{app:induced}

In the induced vacuum, there is a universal matter condition
\begin{equation}
p^M_n|0\rangle_{\mathrm{Ind}}=0
\qquad(\forall n\in\mathbb Z).
\label{eq:ind_vac}
\end{equation}
For each ghost branch, we define the ghost polarization $B_i$ and $C_i$ as
\begin{equation}\label{eq:ghostPolarization}
B_i:\ b_{i,n}|0\rangle_{\mathrm{Ind}}=0,
\qquad
C_i:\ c^i_n|0\rangle_{\mathrm{Ind}}=0
\qquad(\forall n\in\mathbb Z).
\end{equation}
The ILST null, Carroll--Weyl gauged, and hybrid null strings use \(B_i\) on every branch.  In the conformal string, however, \(Q_{\mathrm c}\) contains \(c^1M^1\) and \(M^1|0\rangle_{\mathrm{Ind}}\neq0\), so BRST closure forces us to choose \(C_1\); the complementary coordinate branch takes \(B_{-1}\).  The \(0\) and \(L\) branches remain free, giving  four candidates
\begin{equation}
BBCB,\qquad BBCC,\qquad BCCB,\qquad BCCC,
\label{eq:modelI_induced_polarizations}
\end{equation}
in the order \((-1,0,1,L)\).

Let \(\rho_i=0\) for \(B_i\) and \(\rho_i=1\) for \(C_i\).  The non-zero contractions are fixed algebraically by the canonical brackets and vacuum definition:
\begin{align}
\langle p^N_lx^M_k\rangle_{\mathrm{Ind}}
&=-i\eta^{MN}\delta_{k+l,0},
&
\langle x^M_kp^N_l\rangle_{\mathrm{Ind}}&=0,
\label{eq:ind_px}\\
\langle c^i_kb_{j,l}\rangle_{\mathrm{Ind}}
&=\rho_i\delta^i_j\delta_{k+l,0},
&
\langle b_{j,l}c^i_k\rangle_{\mathrm{Ind}}
&=(1-\rho_i)\delta^i_j\delta_{k+l,0}.
\label{eq:ind_cb}
\end{align}

Following the generalized-zeta family of Ref.~\cite{Hwang:1998brst}, we define
\begin{equation}
S_\alpha\equiv
\left.\sum_{r\in\mathbb Z}1\right|_\alpha=-2\alpha.
\label{eq:Salpha_induced}
\end{equation}
For the bilinear matter generators, every \(xp\)--\(xp\) double contraction contains a vanishing \(\langle xp\rangle\).  The same argument works for the all-\(B\) ghost polarization in the ILST null, Carroll--Weyl gauged, and hybrid null strings.  The only constraint-algebra exception is the conformal string worldsheet special-conformal--supertranslation pair \(M^1\sim xx\), \(M^{-1}\sim pp\).  For a fixed ordered sector pair \((i,j)\), \(\mathcal K^{ij}_{\rm mat}(m)\) and \(\mathcal K^{ij}_{\rm gh}(m)\) denote the matter and ghost double-contraction contributions to the unshifted central coefficient \(K^{ij}_m\), before the \(A_i\) shifts and before imposing \(\alpha=0\); their sum is \(\mathcal K^{ij}(m)\).  In the conformal string sector displayed next, these pre-regularization contributions are independent of \(m\).  The matter and ghost contributions are
\begin{align}
\mathcal K^{1,-1}_{\rm mat}(m)
&=\frac{D+2}{2}S_\alpha,
&
\mathcal K^{1,-1}_{\rm gh}(m)
&=2(\rho_{-1}-\rho_1)S_\alpha=-2S_\alpha.
\label{eq:app_modelI_ghost_induced}
\end{align}
All four polarizations in Eq.~\eqref{eq:modelI_induced_polarizations} therefore give
\begin{align}
\mathcal K^{L,L}(m)
&=\mathcal K^{0,L}(m)
=\mathcal K^{0,0}(m)=0,
&
\mathcal K^{1,-1}(m)
&=\frac{D-2}{2}S_\alpha.
\label{eq:app_modelI_induced_bare}
\end{align}
After the \(A_0,A_L\) zero-mode shifts,
\begin{align}
d^{L,L}_m&=2mA_L,
&
d^{0,L}_m&=mA_0,
&
d^{0,0}_m&=0,
&
d^{1,-1}_m&=\frac{D-2}{2}S_\alpha+2iA_0.
\end{align}
The conformal string Jacobi identities require the same \(\tilde d_4\) in \(d^{0,L}_m=\tilde d_4m\) and \(d^{1,-1}_m=2i\tilde d_4\) when \(\tilde d_3=0\).  Their shift-independent mismatch is
\begin{equation}
\mathcal J_m
=d^{1,-1}_m-\frac{2i}{m}d^{0,L}_m
=\frac{D-2}{2}S_\alpha,
\qquad m\neq0.
\label{eq:app_modelI_induced_J}
\end{equation}
A non-zero \(\mathcal J_m\) is not a two-cocycle and is therefore not interpreted as a critical-dimension condition.  We instead impose the symmetric member of the regulator family in Ref.~\cite{Hwang:1998brst},
\begin{equation}
\alpha=0
\quad\Longrightarrow\quad
S_0=1+2\zeta(0)=0,
\end{equation}
which restores the Jacobi identity for arbitrary \(D\).

For completeness, the full induced-vacuum coefficient set is collected below.  The middle column contains shift-independent non-trivial or exceptional coefficients; the last column contains the removable coefficients in the representative \(A_{-1}=0\).
\begin{center}
\resizebox{\textwidth}{!}{%
\begin{tabular}{lcc}
\toprule
String & Computed shift-independent coefficients & Removable coefficients\\
\midrule
Conformal string & \(\tilde d_1=\tilde d_3=0\)
  & \(\tilde d_2=2A_L,\ \tilde d_4=A_0\)\\
ILST null string & \(\tilde d_1=\tilde d_5=0\)
   & \(\tilde d_2=2A_L,\ \tilde d_6=2A_{-1}=0\)\\
Carroll--Weyl gauged string & \(\tilde d_1=\tilde d_3=\tilde d'_3=0\)
    & \(\tilde d_2=2A_L,\ \tilde d_4=A_0\)\\
Hybrid null string & \(\tilde d_1=0;\ \text{all exceptional coefficients}=0\)
   & \(\tilde d_2=2A_{L_\lambda},\
      g_\lambda^{\rm lin}(m)=(\lambda+1)A_{-1}m=0\)\\
\bottomrule
\end{tabular}
}
\end{center}
For the hybrid null string, the \emph{exceptional central extensions} are the constant and linear terms at \(\lambda=-1\), the quadratic term and \(h_0(m)\) at \(\lambda=0\), and the cubic term at \(\lambda=1\); all evaluate to zero.  The separately displayed \(g_\lambda^{\rm lin}(m)\) is the removable linear piece and vanishes in the chosen \(A_{-1}=0\) representative.  Thus the induced prescription selects no critical dimension in any of the four models.

\section{Derivation of worldsheet quantum anomaly in the flipped vacuum}
\label{app:flipped_derivations}

The flipped vacuum is defined by the conditions
\begin{align}
x^M_n|0\rangle_{\rm Flip}
=p^M_n|0\rangle_{\rm Flip}
=b_{i,n}|0\rangle_{\rm Flip}
=c^i_n|0\rangle_{\rm Flip}&=0
\quad(n>0),\nonumber\\
p^M_0|0\rangle_{\rm Flip}
=b_{i,0}|0\rangle_{\rm Flip}&=0.
\end{align}
The corresponding non-zero two-point functions are
\begin{align}
\langle x^M_kp^N_l\rangle_{\rm Flip}
&=i\eta^{MN}\delta_{k+l,0}\Theta(k>0),
&
\langle p^N_lx^M_k\rangle_{\rm Flip}
&=-i\eta^{MN}\delta_{k+l,0}\Theta(l\geq0),\nonumber\\
\langle b_{i,k}c^j_l\rangle_{\rm Flip}
&=\delta_i^j\delta_{k+l,0}\Theta(k\geq0),
&
\langle c^j_lb_{i,k}\rangle_{\rm Flip}
&=\delta_i^j\delta_{k+l,0}\Theta(l>0).
\label{eq:flipped_2pt}
\end{align}
Here \(\Theta(\mathsf C)=1\) when the condition \(\mathsf C\) is true and \(0\) otherwise.  The strict versus non-strict inequalities implement the prescribed zero-mode ordering: \(b_{i,0}\) annihilates the ket and is placed to the right of \(c^i_0\), while no annihilation condition is imposed on \(c^i_0\) or \(x^M_0\).
For \(m>0\), the shift term vanishes and the positive-frequency constraint is \(\tilde\phi^i_m=H^i_m=M^i_m+G^i_m\). Since every such constraint annihilates the ket, all central coefficients follow from the single relation
\begin{equation}
d^{ij}_m
=\langle M^i_mM^j_{-m}\rangle_{\rm Flip}
+\langle G^i_mG^j_{-m}\rangle_{\rm Flip}
+U^{ij}{}_{r}(m,-m)A_r.
\label{eq:flipped_master}
\end{equation}
The step functions restrict every non-zero bilinear double contraction to a finite range.
We now work out the conformal string in detail and use the same method to state the remaining models without repeating identical contraction bookkeeping.

\subsection{The conformal string}\label{app:modelI}

For the Virasoro pair, the matter trace runs over \(D+2\) target coordinates and the four ghost branches give
\begin{align}
\langle M^L_mM^L_{-m}\rangle_{\rm Flip}
&=(D+2)\frac{m^3-m}{6},&
\langle G^L_{{\mathrm c},m}G^L_{{\mathrm c},-m}\rangle_{\rm Flip}
&=-\frac{14}{3}m^3+\frac23m.
\end{align}
Equation~\eqref{eq:flipped_master} therefore yields
\begin{equation}
d^{L,L}_m
=\frac{D-26}{6}m^3
+\left(2A_L-\frac{D-2}{6}\right)m,
\qquad
\tilde d_1=\frac{D-26}{6},\quad
\tilde d_2=2A_L-\frac{D-2}{6}.
\label{eq:flip_modelI_LL}
\end{equation}
For the scaling--Virasoro pair,
\begin{align}
\langle M^0_mM^L_{-m}\rangle_{\rm Flip}
&=i\frac{D+2}{4}(m^2-m),&
\langle G^0_{{\mathrm c},m}G^L_{{\mathrm c},-m}\rangle_{\rm Flip}
&=-2im^2,
\end{align}
and hence
\begin{equation}
d^{0,L}_m
=i\left(\frac{D+2}{4}-2\right)m^2
+\left(A_0-i\frac{D+2}{4}\right)m.
\end{equation}
The scaling--scaling contractions give
\begin{equation}
    d^{0,0}_m=\left(\frac{D+2}{4}-2\right)m\,.
\end{equation}
The contraction between the worldsheet special-conformal generators and the supertranslation generators independently gives
\begin{align}
\langle M^1_mM^{-1}_{-m}\rangle_{\rm Flip}
&=-\frac{D+2}{2}(m-1),&
\langle G^1_{{\mathrm c},m}G^{-1}_{{\mathrm c},-m}\rangle_{\rm Flip}
&=4m.
\end{align}
Hence, we have
\begin{equation}
    d^{1,-1}_m=-\frac{D-6}{2}m+\frac{D+2}{2}+2iA_0.
\end{equation}
All three sectors agree with the Jacobi-linked parameters
\begin{equation}
\tilde d_3=i\left(\frac{D+2}{4}-2\right),
\qquad
\tilde d_4=A_0-i\frac{D+2}{4}.
\label{eq:flip_modelI_scaling}
\end{equation}
Thus \(\tilde d_1=0\) requires \(D=26\), whereas \(\tilde d_3=0\) requires \(D=6\); the conformal string has no consistent flipped-vacuum dimension.

\subsection{The ILST null string}\label{app:modelII}

The only non-zero central sector is Virasoro--Virasoro:
\begin{align}
\langle M^L_mM^L_{-m}\rangle_{\rm Flip}
&=D\frac{m^3-m}{6},&
\langle G^L_{{\mathrm n},m}G^L_{{\mathrm n},-m}\rangle_{\rm Flip}
&=-\frac{13}{3}m^3+\frac13m,\nonumber\\
d^{L,L}_m
&=\frac{D-26}{6}m^3
+\left(2A_L-\frac{D-2}{6}\right)m.
\end{align}
Consequently,
\begin{equation}
\tilde d_1=\frac{D-26}{6},
\qquad
\tilde d_2=2A_L-\frac{D-2}{6},
\qquad
\tilde d_5=\tilde d_6=0.
\end{equation}
The genuine anomaly vanishes at \(D=26\).

\subsection{The Carroll--Weyl gauged string}\label{app:modelIII}

The three ghost branches modify the Virasoro sum to
\begin{align}
\langle M^L_mM^L_{-m}\rangle_{\rm Flip}
&=D\frac{m^3-m}{6},&
\langle G^L_{{\mathrm g},m}G^L_{{\mathrm g},-m}\rangle_{\rm Flip}
&=-\frac92m^3+\frac12m,
\end{align}
so that
\begin{equation}
\tilde d_1=\frac{D-27}{6},
\qquad
\tilde d_2=2A_L-\frac{D-3}{6}.
\end{equation}
The remaining non-zero contractions are
\begin{align}
\langle M^0_mM^L_{-m}\rangle_{\rm Flip}
&=i\frac D4(m^2-m),&
\langle G^0_{{\mathrm g},m}G^L_{{\mathrm g},-m}\rangle_{\rm Flip}
&=-\frac{3i}{2}m^2+\frac i2m,
\end{align}
so that
\begin{equation}
    d^{0,0}_m=\left(\frac D4-1\right)m.
\end{equation}
In conclusion, we have
\begin{equation}
\tilde d'_3=i\left(\frac D4-\frac32\right),
\qquad
\tilde d_3=i\left(\frac D4-1\right),
\qquad
\tilde d_4=A_0-i\left(\frac D4-\frac12\right).
\end{equation}
The conditions are \(D=27\) from \(\tilde d_1\), \(D=6\) from \(\tilde d'_3\), and \(D=4\) from \(\tilde d_3\); they cannot hold simultaneously.

\subsection{The hybrid null string}\label{app:modelIV}

For \(M^{L_\lambda}\), the matter and two-branch ghost contractions are
\begin{align}
\langle M^{L_\lambda}_mM^{L_\lambda}_{-m}\rangle_{\rm Flip}
&=\frac{D(3\lambda^2-1)}{12}m^3-\frac D6m,\nonumber\\
\langle G^{L_\lambda}_{{\mathrm h},m}G^{L_\lambda}_{{\mathrm h},-m}\rangle_{\rm Flip}
&=-\left(\lambda^2+\lambda+\frac73\right)m^3+\frac13m.
\end{align}
Therefore
\begin{equation}
\tilde d_1
=\frac{D(3\lambda^2-1)-2(6\lambda^2+6\lambda+14)}{12},
\qquad
\tilde d_2=2A_{L_\lambda}-\frac{D-2}{6}.
\label{eq:flip_modelIV}
\end{equation}
The Virasoro anomaly vanishes at the \(D(\lambda)\) stated in Section~\ref{sec:anomalies}.

Finally, all central extensions involved with the supertranslation constraint $\tilde{\phi}^{-1}_m$ not displayed above vanish for the same two reasons: the matter sector would require a non-zero \(\langle pp\rangle\), and the ghost sector would require contractions between different branches.  This proves \(\tilde d_5=\tilde d_6=0\) in the ILST null string and gives no independent worldsheet-supertranslation central terms in the conformal and Carroll--Weyl gauged strings.  In the hybrid null string it also sets to zero every Jacobi-allowed exceptional central extension $g_\lambda(m)$ and $h_\lambda(m)$. 
\section{Spacetime conformal symmetry in the ILST null string}
\label{app:SO(D,2)}

Let \(b^\mu\) be a constant target-space parameter, unrelated to the antighost modes \(b_{i,n}\).  The covariant ILST action is invariant under a target-space special conformal transformation
\begin{equation}
\delta_bX^\mu=(b\cdot X)X^\mu-\frac12X^2b^\mu,
\qquad
\delta_bV^{\alpha}=-(b\cdot X)V^{\alpha}.
\label{eq:SCT_transformation}
\end{equation}
To distinguish global target-space charges from the local momentum field \(P^\mu(\sigma)\) and from the worldsheet constraints \(M^i_m\), we define spacetime momentum and angular momentum as
\begin{align}
\mathsf P^\mu
&\equiv\int_0^{2\pi}d\sigma\,P^\mu(\sigma)=p^\mu_0,\nonumber\\
\mathsf M^{\mu\nu}
&\equiv\int_0^{2\pi}d\sigma\,
\bigl(X^\mu P^\nu-X^\nu P^\mu\bigr)
=\sum_{n\in\mathbb Z}\mathopen{:}
\bigl(x^\mu_{-n}p^\nu_n-x^\nu_{-n}p^\mu_n\bigr)
\mathclose{:}.
\end{align}
The matter-sector contribution to the associated conserved charge is
\begin{align}
K^\mu_{\rm matter}
&=\int_0^{2\pi}d\sigma\,
\left[(X\cdot P)X^\mu-\frac12X^2P^\mu\right]\nonumber\\
&=\sum_{l,k}\mathopen{:}
\left[(x_l\cdot p_k)x^\mu_{-l-k}
-\frac12(x_l\cdot x_k)p^\mu_{-l-k}\right]\mathclose{:}.
\label{eq:K_matter}
\end{align}
The bare matter charge $K^\mu_\text{matter}$ does not commute with the ILST null string BRST charge $Q_\mathrm n$. We can determine the ghost completion without repeating the gauge-fixing analysis by requiring that the completed conserved charge classically commutes with the BRST charge.  

The completed dilation and special-conformal generators are
\begin{align}
\mathcal D&=\mathcal D_{\rm matter}+\mathcal D_{\rm ghost},
&
\mathcal D_{\rm matter}&=\sum_n\mathopen{:}x_{-n}\cdot p_n\mathclose{:},
&
\mathcal D_{\rm ghost}&=-2iJ_0,
\label{eq:D_ghost}\\
K^\mu&=K^\mu_{\rm matter}+K^\mu_{\rm ghost},
&
K^\mu_{\rm ghost}
&=-2i\sum_{l,k}\mathopen{:}
x^\mu_{-l}J_{l}\mathclose{:}.
\label{eq:K_ghost}
\end{align}
with
\begin{equation}
J_m=\sum_k\mathopen{:}c^{-1}_k b_{-1,m-k}\mathclose{:}.
\end{equation}
Here only the \((-1)\) ghost branch appears, since \(M^{-1}\) is the only ILST null string constraint carrying target-space dilation weight. 

One can directly check that the single-contraction evaluation gives
\begin{equation}
[\mathcal D,Q_{\mathrm n,0}]_{\rm classical}=0,
\qquad
[K^\mu,Q_{\mathrm n,0}]_{\rm classical}=0.
\label{eq:K_full_BRST}
\end{equation}
The removable shift \(Q_{\mathrm n}-Q_{\mathrm n,0}=-A_Lc^L_0\) does not affect these relations because the displayed charges commute with \(c^L_0\).
Thus, the displayed commutators also hold for $Q_\mathrm n$ with arbitrary \(A_L\).

\subsection{Induced-vacuum}

In this subsection, every colon in the composite charges above is evaluated with induced-vacuum normal ordering.
In the induced vacuum of the ILST null string, we choose all-$B$ ghost polarization \eqref{eq:ghostPolarization}, \(p_n,b_{-1,n},b_{L,n}\) annihilate the induced ket state for every \(n\).  Thus \(\langle px\rangle\) and \(\langle bc\rangle\) are non-zero, whereas the reversed elementary contractions vanish.  All double contractions in \([K^\mu,Q_{\mathrm n}]\) vanish except the reversed contraction of the two momenta in \(c^{-1}M^{-1}\) with the two coordinates in \(K^\mu_{\rm matter}\).
Using the common contact trace \(S_\alpha\) defined in Eq.~\eqref{eq:Salpha_induced}, before imposing the \(\alpha=0\) prescription,
\begin{equation}
\mathcal A^\mu_{\mathrm{Ind}}
\equiv\langle\!\langle[K^\mu,Q_{\mathrm n}]\rangle\!\rangle_{\mathrm{Ind}}
=-\frac{D-2}{2}S_\alpha
\sum_n c^{-1}_{-n}p^\mu_n.
\label{eq:KQ_induced_correction}
\end{equation}
For \(S_\alpha\neq0\) and \(D\neq2\), the non-zero displayed operator is not by itself \(Q_{\mathrm n}\)-closed.  At \(D=2\) its coefficient vanishes algebraically. However, we do not take $D=2$ as the critical dimension. We adopt the symmetric prescription used throughout this paper, which is
\begin{equation}
\alpha=0,
\qquad
S_0=1+2\zeta(0)=0.
\end{equation}
It removes the obstruction for arbitrary \(D\).  The remaining target-space charges \(\mathsf P^\mu\), \(\mathsf M^{\mu\nu}\), \(\mathcal D\), and \(K^\mu\) then obey the standard \(SO(D,2)\) brackets by canonical single contractions and are \(Q_{\mathrm n}\)-closed.  This is a statement about the operator algebra with suitable regularization, not a construction of a positive-definite physical Hilbert space.

\subsection{Flipped-vacuum}

In this subsection, every normal ordered conserved charge is defined associated with flipped-vacuum.  With the two-point functions in Eq.~\eqref{eq:flipped_2pt}, the double-contraction part is
\[
\begin{aligned}
\mathcal A^\mu_{\rm Flip}
&\equiv\langle\!\langle[K^\mu,Q_{\mathrm n}]\rangle\!\rangle_{\rm Flip}\\
&=\langle\!\langle[K^\mu_{\rm matter},
Q_{{\mathrm n},0,{\rm matter}}]\rangle\!\rangle_{\rm Flip}
+\langle\!\langle[K^\mu_{\rm ghost},
Q_{{\mathrm n},0,{\rm matter}}]\rangle\!\rangle_{\rm Flip}\\
&\quad+\langle\!\langle[K^\mu_{\rm ghost},
Q_{{\mathrm n},0,{\rm ghost}}]\rangle\!\rangle_{\rm Flip}.
\end{aligned}
\]
This quantity does not vanish.  The matter charge commutes with the ghost part of \(Q_{\mathrm n}\), and all branch-mismatched or \(pp\) contractions vanish.  The surviving terms exhaust the two possible uncontracted operator classes:
\begin{center}
\begin{tabular}{lcc}
\toprule
source & coefficient of \(c^L_{-n}x^\mu_n\) & coefficient of \(c^{-1}_{-n}p^\mu_n\)\\
\midrule
matter--matter & \(iD\,n(n+1)/2\) & \(-\frac{D-2}{2}(n+1)\)\\
ghost--ghost & \(-i(3n^2+n)\) & \(0\)\\
ghost--matter & \(0\) & \(2n\)\\
\bottomrule
\end{tabular}
\end{center}
The entries follow from the two inequivalent Wick pairings and the finite sums selected by the step functions; the two orderings combine into a single sum for each operator class.  Their sum is
\begin{equation}
\boxed{
\begin{aligned}
\mathcal A^\mu_{\rm Flip}
&=i\sum_{n\neq0}
\left[D\frac{n(n+1)}{2}-(3n^2+n)\right]
c^L_{-n}x^\mu_n\\
&\quad+\sum_n
\left[-\frac{D-2}{2}(n+1)+2n\right]
c^{-1}_{-n}p^\mu_n .
\end{aligned}}
\label{eq:A_mu_explicit}
\end{equation}
Hence no value of \(D\) makes the fixed displayed \(K^\mu\) BRST closed.  In particular, at \(D=26\) the zero-mode-shifted worldsheet charge \(Q_{\mathrm n}\) can be nilpotent while the target-space special-conformal representative remains anomalous.
\bibliographystyle{JHEP}
\bibliography{biblio.bib}

@article{Isberg:1993av,
    author = "Isberg, J. and Lindstrom, U. and Sundborg, B. and Theodoridis, G.",
    title = "{Classical and quantized tensionless strings}",
    eprint = "hep-th/9307108",
    archivePrefix = "arXiv",
    reportNumber = "USITP-93-12",
    doi = "10.1016/0550-3213(94)90056-6",
    journal = "Nucl. Phys. B",
    volume = "411",
    pages = "122--156",
    year = "1994"
}

@article{Gustafsson:1994kr,
    author = "Gustafsson, H. and Lindstrom, U. and Saltsidis, P. and Sundborg, B. and van Unge, R.",
    title = "{Hamiltonian BRST quantization of the conformal string}",
    eprint = "hep-th/9410143",
    archivePrefix = "arXiv",
    reportNumber = "USITP-94-08",
    doi = "10.1016/0550-3213(95)00051-S",
    journal = "Nucl. Phys. B",
    volume = "440",
    pages = "495--520",
    year = "1995"
}

@article{Hwang:1998brst,
    author = "Hwang, Stephen and Marnelius, Robert and Saltsidis, Panagiotis",
    title = "{A general BRST approach to string theories with zeta function regularizations}",
    eprint = "hep-th/9804003",
    archivePrefix = "arXiv",
    reportNumber = "ITP-Goteborg 98-2, DAMTP-98-21, HKS-NT-FR-98/2-SE",
    doi = "10.1063/1.532994",
    journal = "J. Math. Phys.",
    volume = "40",
    number = "10",
    pages = "4639--4657",
    year = "1999"
}

@article{Sheikh-Jabbari:2026vqh,
    author = "Sheikh-Jabbari, M. M. and Yavartanoo, H.",
    title = "{Null Strings Gauged and Reloaded, I: Null Strings Have Carroll-Weyl Gauge Symmetry}",
    eprint = "2605.25817",
    archivePrefix = "arXiv",
    primaryClass = "hep-th",
    month = "5",
    year = "2026"
}

@article{Sheikh-Jabbari:2026tpf,
    author = "Sheikh-Jabbari, M. M. and Yavartanoo, H.",
    title = "{Null Strings Gauged and Reloaded, II: Consistent Classical Treatment of the Null Strings}",
    eprint = "2605.26822",
    archivePrefix = "arXiv",
    primaryClass = "hep-th",
    month = "5",
    year = "2026"
}

@article{Chen:2026klv,
    author = "Chen, Bin and Hu, Zezhou",
    title = "{Symmetries and Critical Dimensions of Tensionless Branes}",
    eprint = "2604.01883",
    archivePrefix = "arXiv",
    primaryClass = "hep-th",
    month = "4",
    year = "2026"
}

@article{Atick:1988si,
    author = "Atick, Joseph J. and Witten, Edward",
    title = "{The Hagedorn Transition and the Number of Degrees of Freedom of String Theory}",
    reportNumber = "IASSNS-HEP-88-14",
    doi = "10.1016/0550-3213(88)90151-4",
    journal = "Nucl. Phys. B",
    volume = "310",
    pages = "291--334",
    year = "1988"
}

@article{Schild:1977vl,
    author = "Schild, Alfred",
    title = "{Classical Null Strings}",
    reportNumber = "PRINT-76-0491 (TEXAS), ANL-HEP-PR-77-23",
    doi = "10.1103/PhysRevD.16.1722",
    journal = "Phys. Rev. D",
    volume = "16",
    pages = "1722",
    year = "1977"
}

@article{Karlhede:1986pq,
    author = "Karlhede, A. and Lindstrom, U.",
    title = "{The Classical Bosonic String in the Zero Tension Limit}",
    doi = "10.1088/0264-9381/3/4/002",
    journal = "Class. Quant. Grav.",
    volume = "3",
    pages = "L73--L75",
    year = "1986"
}

@article{Sundborg:1994aa,
    author = "Sundborg, Bo",
    title = "{Strongly topological interactions of tensionless strings}",
    eprint = "hep-th/9405195",
    archivePrefix = "arXiv",
    reportNumber = "USITP-94-04",
    month = "5",
    year = "1994"
}

@article{Sundborg:2000wp,
    author = "Sundborg, Bo",
    editor = "Sorokin, Dmitri P.",
    title = "{Stringy gravity, interacting tensionless strings and massless higher spins}",
    eprint = "hep-th/0103247",
    archivePrefix = "arXiv",
    doi = "10.1016/S0920-5632(01)01545-6",
    journal = "Nucl. Phys. B Proc. Suppl.",
    volume = "102",
    pages = "113--119",
    year = "2001"
}

@article{Gross:1987ar,
    author = "Gross, David J. and Mende, Paul F.",
    title = "{The High-Energy Behavior of String Scattering Amplitudes}",
    reportNumber = "PUPT-1062",
    doi = "10.1016/0370-2693(87)90355-8",
    journal = "Phys. Lett. B",
    volume = "197",
    pages = "129--134",
    year = "1987"
}

@article{Gross:1988ue,
    author = "Gross, David J. and Mende, Paul F.",
    title = "{String Theory Beyond the Planck Scale}",
    reportNumber = "PUPT-1067",
    doi = "10.1016/0550-3213(88)90390-2",
    journal = "Nucl. Phys. B",
    volume = "303",
    pages = "407--454",
    year = "1988"
}

@article{Barnich:2010eb,
    author = "Barnich, Glenn and Troessaert, Cedric",
    title = "{Symmetries of asymptotically flat 4 dimensional spacetimes at null infinity revisited}",
    eprint = "0909.2617",
    archivePrefix = "arXiv",
    primaryClass = "gr-qc",
    reportNumber = "ULB-TH-09-24",
    doi = "10.1103/PhysRevLett.105.111103",
    journal = "Phys. Rev. Lett.",
    volume = "105",
    pages = "111103",
    year = "2010"
}

@article{Bagchi:2013bga,
    author = "Bagchi, Arjun",
    title = "{Tensionless Strings and Galilean Conformal Algebra}",
    eprint = "1303.0291",
    archivePrefix = "arXiv",
    primaryClass = "hep-th",
    reportNumber = "MIT-CTP-4445, EMPG-13-02",
    doi = "10.1007/JHEP05(2013)141",
    journal = "JHEP",
    volume = "05",
    pages = "141",
    year = "2013"
}

@article{Ciambelli:2019lap,
    author = "Ciambelli, Luca and Leigh, Robert G. and Marteau, Charles and Petropoulos, P. Marios",
    title = "{Carroll Structures, Null Geometry and Conformal Isometries}",
    eprint = "1905.02221",
    archivePrefix = "arXiv",
    primaryClass = "hep-th",
    reportNumber = "CPHT-RR025.052019, CPHT-RR010.022019",
    doi = "10.1103/PhysRevD.100.046010",
    journal = "Phys. Rev. D",
    volume = "100",
    number = "4",
    pages = "046010",
    year = "2019"
}

@article{Barnich:2012aw,
    author = "Barnich, Glenn and Gomberoff, Andres and Gonzalez, Hernan A.",
    title = "{The Flat limit of three dimensional asymptotically anti-de Sitter spacetimes}",
    eprint = "1204.3288",
    archivePrefix = "arXiv",
    primaryClass = "gr-qc",
    doi = "10.1103/PhysRevD.86.024020",
    journal = "Phys. Rev. D",
    volume = "86",
    pages = "024020",
    year = "2012"
}

@article{Bagchi:2012cy,
    author = "Bagchi, Arjun and Fareghbal, Reza",
    title = "{BMS/GCA Redux: Towards Flatspace Holography from Non-Relativistic Symmetries}",
    eprint = "1203.5795",
    archivePrefix = "arXiv",
    primaryClass = "hep-th",
    reportNumber = "EMPG-12-04, NI-12014",
    doi = "10.1007/JHEP10(2012)092",
    journal = "JHEP",
    volume = "10",
    pages = "092",
    year = "2012"
}

@article{Bagchi:2012yk,
    author = "Bagchi, Arjun",
    title = "{Correspondence between Asymptotically Flat Spacetimes and Nonrelativistic Conformal Field Theories}",
    eprint = "1006.3354",
    archivePrefix = "arXiv",
    primaryClass = "hep-th",
    doi = "10.1103/PhysRevLett.105.171601",
    journal = "Phys. Rev. Lett.",
    volume = "105",
    pages = "171601",
    year = "2010"
}

@article{Bagchi:2015nca,
    author = "Bagchi, Arjun and Chakrabortty, Shankhadeep and Parekh, Pulastya",
    title = "{Tensionless Strings from Worldsheet Symmetries}",
    eprint = "1507.04361",
    archivePrefix = "arXiv",
    primaryClass = "hep-th",
    reportNumber = "MIT-CTP-4690",
    doi = "10.1007/JHEP01(2016)158",
    journal = "JHEP",
    volume = "01",
    pages = "158",
    year = "2016"
}

@article{Bagchi:2020zmr,
    author = "Bagchi, Arjun and Banerjee, Aritra and Parekh, Pulastya",
    title = "{Tensionless Path from Closed to Open Strings}",
    eprint = "1905.11732",
    archivePrefix = "arXiv",
    primaryClass = "hep-th",
    doi = "10.1103/PhysRevLett.123.111601",
    journal = "Phys. Rev. Lett.",
    volume = "123",
    number = "11",
    pages = "111601",
    year = "2019"
}

@article{Bagchi:2021ban,
    author = "Bagchi, Arjun and Banerjee, Aritra and Chakrabortty, Shankhadeep and Chatterjee, Ritankar",
    title = "{A Rindler road to Carrollian worldsheets}",
    eprint = "2111.01172",
    archivePrefix = "arXiv",
    primaryClass = "hep-th",
    doi = "10.1007/JHEP04(2022)082",
    journal = "JHEP",
    volume = "04",
    pages = "082",
    year = "2022"
}

@article{Bagchi:2021pci,
    author = "Bagchi, Arjun and Mandlik, Mangesh and Sharma, Punit",
    title = "{Tensionless tales: vacua and critical dimensions}",
    eprint = "2105.09682",
    archivePrefix = "arXiv",
    primaryClass = "hep-th",
    doi = "10.1007/JHEP08(2021)054",
    journal = "JHEP",
    volume = "08",
    pages = "054",
    year = "2021"
}

@article{Strominger:2013jfa,
    author = "Strominger, Andrew",
    title = "{On BMS Invariance of Gravitational Scattering}",
    eprint = "1312.2229",
    archivePrefix = "arXiv",
    primaryClass = "hep-th",
    doi = "10.1007/JHEP07(2014)152",
    journal = "JHEP",
    volume = "07",
    pages = "152",
    year = "2014"
}

@article{Donnay:2022aba,
    author = "Donnay, Laura and Fiorucci, Adrien and Herfray, Yannick and Ruzziconi, Romain",
    title = "{Carrollian Perspective on Celestial Holography}",
    eprint = "2202.04702",
    archivePrefix = "arXiv",
    primaryClass = "hep-th",
    doi = "10.1103/PhysRevLett.129.071602",
    journal = "Phys. Rev. Lett.",
    volume = "129",
    number = "7",
    pages = "071602",
    year = "2022"
}

@article{Duary:2026rlo,
    author = "Duary, Sarthak and Maji, Sourav",
    title = "{Path integral quantization of null bosonic strings with Carroll-Weyl ghosts}",
    eprint = "2606.04999",
    archivePrefix = "arXiv",
    primaryClass = "hep-th",
    month = "6",
    year = "2026"
}

@article{Sheikh-Jabbari:2026cnj,
    author = "Sheikh-Jabbari, M. M. and Yavartanoo, H.",
    title = "{On the Consistency of Null Strings Literature: The Tale of an Overlooked Symmetry}",
    eprint = "2605.12414",
    archivePrefix = "arXiv",
    primaryClass = "hep-th",
    month = "5",
    year = "2026"
}

@article{Lindstrom:2026quz,
    author = {Lindstr{\"o}m, Ulf},
    title = "{Symmetries of tensionless strings}",
    eprint = "2605.26185",
    archivePrefix = "arXiv",
    primaryClass = "hep-th",
    reportNumber = "Uppsala Institute for Theoretical Physics preprint UUITP-08/26",
    month = "5",
    year = "2026"
}

@article{Lindstrom:2026zno,
    author = {Lindstr{\"o}m, Ulf},
    title = "{The conformal null string in $d+2$ and $d$ dimensions}",
    eprint = "2606.22498",
    archivePrefix = "arXiv",
    primaryClass = "hep-th",
    reportNumber = "Uppsala University, Theoretical Physics UUITP-15/26",
    month = "6",
    year = "2026"
}

@article{Figueroa-OFarrill:2024wgs,
    author = "Figueroa-O'Farrill, Jos{\'e} M. and Vishwa, Girish S.",
    title = "{The BRST quantisation of chiral BMS-like field theories}",
    eprint = "2407.12778",
    archivePrefix = "arXiv",
    primaryClass = "hep-th",
    reportNumber = "EMPG-24-3",
    doi = "10.1063/5.0237868",
    journal = "J. Math. Phys.",
    volume = "66",
    number = "4",
    pages = "042303",
    year = "2025"
}

@article{Gao:2011Wab,
    author = "Gao, Shoulan and Jiang, Cuipo and Pei, Yufeng",
    title = "{Low-Dimensional Cohomology Groups of the Lie Algebras $W(a,b)$}",
    doi = "10.1080/00927871003591835",
    journal = "Commun. Algebra",
    volume = "39",
    number = "2",
    pages = "397--423",
    year = "2011"
}

@article{Rasulian:2026lc,
    author = "Rasulian, Ida M. and Sheikh-Jabbari, M. M. and Yavartanoo, H.",
    title = "{Null-strings Gauged, Reloaded and Quantized, I: Canonical Quantization in the Light-Cone Gauge}",
    eprint = "2607.02970",
    archivePrefix = "arXiv",
    primaryClass = "hep-th",
    month = "7",
    year = "2026"
}

@article{Duary:2026cw,
    author = "Duary, Sarthak and Maji, Sourav",
    title = "{BRST quantization of Carroll-Weyl gauged null strings}",
    eprint = "2608.02731",
    archivePrefix = "arXiv",
    primaryClass = "hep-th",
    month = "8",
    year = "2026"
}

\end{document}